\documentclass[11pt,a4paper]{article}
\pdfoutput=1
\usepackage{amssymb,amsmath,amsfonts}
\usepackage{epsfig}
\usepackage{jheppub}
\usepackage{slashed}
\usepackage[english]{babel}
\usepackage[latin1]{inputenc}
\usepackage{pgf}
\usepackage{graphicx}

\usepackage{amsmath,amssymb,dsfont}
\usepackage{verbatim}
\usepackage{color}

\usepackage[normal] {subfigure}
    \usepackage{multirow}
    \usepackage{tikz}
    \usepackage{boxedminipage}
    \usepackage[latin1]{inputenc}
    \usepackage[T1]{fontenc}
    \usepackage{graphicx}   
    \usepackage{amsfonts}   
    \usepackage{hyperref}   
    \usepackage{mathrsfs}
    \usepackage{fancybox}
\usepackage{hyperref}

\newcommand{\CN}{\mathcal{N}}

\newcommand{\CM}{\mathcal{M}}

\newcommand{\CH}{\mathcal{H}}

\newcommand{\ads}{\mbox{AdS}}
\newcommand{\cft}{\mbox{CFT}}

\newcommand{\Z}{\mathds{Z}}
\newcommand{\C}{\mathds{C}}
\newcommand{\R}{\mathds{R}}
\newcommand{\sphere}{\mbox{S}}

\newcommand{\ff}{h(\lambda)}
\newcommand{\lp}{\protect{\lambda'}}

\newcommand{\myfigref}[1]{~{Fig.~(\ref{#1})}}
\newcommand{\myref}[1]{~{(\ref{#1})}}
\renewcommand{\eqref}[1]{~{(\ref{#1})}}

\newcommand{\nn}{\nonumber}

\newcommand{\ds}{\displaystyle}

\makeatletter
\def\mr@ignsp#1 {\ifx\:#1\@empty\else #1\expandafter\mr@ignsp\fi}%
\newcommand{\multiref}[1]{\begingroup
\xdef\mr@no@sparg{\expandafter\mr@ignsp#1 \: }%
\def\mr@comma{}%
\@for\mr@refs:=\mr@no@sparg\do{\mr@comma\def\mr@comma{,}\ref{\mr@refs}}%
\endgroup}
\makeatother

\newcommand{\hypref}[2]{\ifx\href\asklfhas #2\else\href{#1}{#2}\fi}

\newcommand{\beq}{\begin{equation}}
\newcommand{\eeq}{\end{equation}}
\def\[{\begin{equation}}
\def\]{\end{equation}}
\newcommand{\be}{\begin{eqnarray}}
\newcommand{\ee}{\end{eqnarray}}

\newcommand{\grp}[1]{\mathrm{#1}}

\newcommand{\grSU}{\grp{SU}}
\newcommand{\grU}{\grp{U}}

\newcommand{\la}{\label}
\newcommand{\mycite}[1]{~{\cite{#1}}}

\subheader{\tiny ITEP-TH-51/11}
\title{Strings in $\ads_4\times \C P^3$: finite size
spectrum vs. Bethe Ansatz}

\author[a]{Davide Astolfi,}
\author[a]{Gianluca Grignani,}
\author[a]{Enrico Ser-Giacomi,}
\author[a,b]{A.V.~Zayakin}

\affiliation[a] {Dipartimento di Fisica, Universit\`a di Perugia,\\
I.N.F.N. Sezione di Perugia,\\ Via Pascoli, I-06123 Perugia,
Italy}
\affiliation[b]{Institute of Theoretical and Experimental Physics,\\
B.~Cheremushkinskaya ul. 25, 117259 Moscow, Russia}

\emailAdd{davide.astolfi@pg.infn.it}
\emailAdd{gianluca.grignani@pg.infn.it}
\emailAdd{erikgs@hotmail.it}
\emailAdd{a.zayakin@gmail.com}

\abstract{We compute the first curvature corrections to the
spectrum of light-cone gauge type IIA string theory that arise in
the expansion of $\ads_4\times \C P^3$ about a plane-wave limit.
The resulting spectrum is shown to match precisely, both in
magnitude and degeneration that of the corresponding solutions of
the all-loop Gromov--Vieira Bethe Ansatz. The one-loop dispersion
relation correction is calculated for all the single oscillator
states of the theory, with the level matching  condition lifted.
It is shown to have all logarithmic divergences cancelled and to
leave only a finite exponentially suppressed contribution, as
shown earlier for light bosons. We argue that there is no
ambiguity in the choice of the regularization for the self-energy
sum, since the regularization applied is the only one preserving
unitarity.   Interaction matrices in the full degenerate
two-oscillator sector are calculated and the spectrum of all two
light magnon oscillators is completely determined. The same
finite-size corrections, at the order $\frac{1}{J}$, where $J$ is
the length of the chain, in the two-magnon sector are calculated
from the all loop Bethe Ansatz. The corrections obtained by the
two completely different methods coincide up to the fourth order
in $\lambda' \equiv \frac{\lambda}{J^2}$. We conjecture that the
equivalence extends to all orders in $\lp$ and to higher orders in
$\frac{1}{J}$. }

\keywords{AdS-CFT correspondence, Penrose Limit and pp-wave background}

\begin{document}
\maketitle
\flushbottom

\setcounter{page}{1}


\section{Introduction}
\label{sec:intro}

The appearance of integrable structures  both at strong and weak
coupling  has given hope for a complete solution to the spectral
problem of the $\ads_5/\cft_4$ duality in the planar limit. This
is the best-understood example of a duality between gauge theory
and string theory, it states the equivalence between the IIB
superstring theory on $\ads_5 \times \sphere^5$ and $\mathcal N=4$
super Yang-Mills (SYM) theory in $3+1$ dimensions.
In~\cite{Beisert:2005fw,Beisert:2006ez} an all-loop asymptotic
Bethe ansatz has been proposed for the $\ads_5/\cft_4$ duality and
starting from the mirror version of the Beisert-Staudacher's
equations, further the Y-system was formulated that alllows
computation of anomalous dimensions for opertaors of any
length~\cite{Gromov:2009tv,Bombardelli:2009ns,Arutyunov:2009ur,Gromov:2009bc}.
This Y-system passes some very important tests: it incorporates
the full asymptotic Beisert-Staudacher's Bethe ansatz at large
length $J$ and it reproduces all known wrapping corrections.

Even if the AdS/CFT correspondence is at present best understood
for $\ads_5$, also in the more recent $\ads_4/\cft_3$ duality the
solution to the spectral problem, at least for the sector
described by a coset space, seems to be at reach in the planar
limit thanks to integrability~\cite{Gromov:2008qe}. The
$\ads_4/\cft_3$ correspondence is an exact duality between type
IIA superstring on $\ads_4 \times \C P^3$ and a certain regime of
the ABJM-theory~\cite{Aharony:2008ug}. The ABJM theory is an
Chern-Simons $\mathcal N=6$  gauge theory  with matter dual to
M-theory compactified onto $\ads_4 \times \sphere^7/\Z_k$. It
possesses a $\grU(N) \times \grU (N )$ gauge symmetry with
Chern-Simons like
 kinetic terms at level $k$ and $-k$; if the 't
Hooft coupling $\lambda=\frac{N}{k}$ is  $1\ll \lambda\ll k^4$ the
gravity side being effectively rendered as a IIA superstring on
$\ads_4 \times \C P^3$.

The  integrability of the $\ads_4/\cft_3$  duality due to the
reduced number of supersymmetries, offers interesting new
challenges. The sector of the theory described by a coset space
was proven to be classically integrable, but classical
integrability for the whole theory has still to be
demonstrated~\cite{Arutyunov:2008if,Stefanski:2008ik,Sorokin:2010wn,Sorokin:2011mj,Cagnazzo:2011at}.
Nevertheless, the semiclassical and quantum integrability of some
sectors of the theory have received plenty of attention both at
weak~\cite{Minahan:2008hf,Gaiotto:2008cg,Bak:2008cp,Kristjansen:2008ib,Zwiebel:2009vb,
Minahan:2009te,Bak:2009tq,Minahan:2009aq,
Minahan:2009wg,LevkovichMaslyuk:2011ty} and at strong coupling
\cite{Grignani:2008is,Grignani:2008te,Astolfi:2008ji,McLoughlin:2008he,
Sundin:2008vt,Zarembo:2009au, Abbott:2011xp}. In particular an
all-loop  asymptotic Bethe Ansatz has been
proposed~\cite{Gromov:2008qe} and a Y-system has been suggested
also for the $\ads_4/\cft_3$
duality~\cite{Gromov:2009tv,Bombardelli:2009ns,
Gromov:2009bc,Bombardelli:2009xz,Gromov:2009at}.
More recently nontrivial evidence for the scattering
amplitudes/Wilson loop duality for this theory has been
given~\cite{Henn:2010ps, Bianchi:2011rn, Chen:2011vv,
Bianchi:2011dg}.

According to the AdS/CFT  correspondence energies of excited
states of superstrings in specific  curved  backgrounds should
coincide with the anomalous dimensions of appropriate operators of
the corresponding gauge field theory. For the planar limit the
coupling is zero, however the string is still in a curved space
and thus its two-dimensional world-sheet theory is not
interaction-free. Calculating the superstring spectrum in such
backgrounds is therefore still a complicated problem. Nevertheless
the nontrivial interactions become small when one takes the
Penrose limit of the
metric~\cite{Metsaev:2001bj,Berenstein:2002jq}. Corrections to the
free spectrum can then be computed perturbatively as an expansion
in inverse powers of the background curvature radius $R$.

This idea was suggested by Callan et al.
in~\cite{Callan:2003xr,Callan:2004uv} for  $\ads_5/\cft_4$
correspondence.  The outcome of~\cite{Callan:2003xr,Callan:2004uv}
and the respective studies on the field theory side
\cite{Beisert:2003tq}, have been important for understanding the
integrability of the AdS/CFT correspondence.
In~\cite{Callan:2003xr,Callan:2004uv} it was shown for the first
time that there is a disagreement  between field theory operators
anomalous dimensions and the respective string energies at three
loops. The disagreement was afterwards interpreted as a breakdown
of a double scaling limit and resolved by including the dressing
factor that interpolates nontrivially from weak to strong coupling
in the Bethe equations describing the spectra of the gauge and the
string theory~\cite{Minahan:2002ve, Arutyunov:2004vx,
Hernandez:2006tk, Beisert:2006ez}.

In \cite{Astolfi:2011ju} a complete  calculation of the curvature
corrections to the pp-wave energy of the two oscillator
non-degenerate bosonic states in the decoupled
$\grSU(2)\times\grSU(2)$ sector of type IIA superstring on $\ads_4
\times \C P^3$ was performed. This study was initiated in
\cite{Astolfi:2008ji} and reexamined in
\cite{Sundin:2008vt,Sundin:2009zu}. In \cite{Astolfi:2009qh} the
interacting Hamiltonian for oscillations in the near plane wave
limit of $\ads_4\times \C P^3$ was calculated. This is a crucial
tool for the computations of this Paper and it is given
in\cite{Astolfi:2011ju} by a perturbative expansion in terms of
$1/R$ powers
\beq \label{eq0} H =H_{2,B}+H_{2,F} +  {1\over R} \left( H_{3,B}+
H_{3,BF}\right) + {1\over R^2} \left( H_{4,B}+ H_{4,F}+
H_{4,BF}\right) +\dots. \eeq
where $R$ is the $\C P^3$ radius. For brevity we shall further
refer to ``third-order Hamiltonian'' $H_3\equiv H_{3,B}+ H_{3,BF}$
and ``fourth-order Hamiltonian'' $H_4=H_{4,B}+ H_{4,F}+ H_{4,BF}$.

The quadratic Hamiltonian term, $H_{2B}+H_{2F}$, is the plane-wave
free Hamiltonian from\cite{Astolfi:2011ju} where  fermionic and
bosonic fields are fully decoupled~\cite{Gaiotto:2008cg,
Nishioka:2008gz, Grignani:2008is,Grignani:2009ny}. A peculiarity
of this theory is that in the pp-wave limit the eight massive
bosons and eight massive fermions have different worldsheet
masses. Four fermions and four bosons are ``heavy'', while the
remaining four fermions and four bosons are ``light'' having a
world sheet mass which is $1/2$ of that of the heavy ones.

The unique feature of the $AdS_4$ case is the presence of cubic
terms $H_3$ in the Hamiltonian~\cite{Astolfi:2008ji}. This yields extra
terms for the matrix element of some arbitrary $|f\rangle$ state
in addition to the expectation value of the quartic Hamiltonian $H_{4}$, $\langle f |H_{4}| f\rangle$:
now the energy correction $\delta E_f ^{(2)}$ looks like

\be \label{eq1} \delta E_f ^{(2)}=\frac{1}{R^2}
\left(\sum_{|i\rangle}\frac{\left|\langle i|H_{3}|f
\rangle\right|^2}{E_f-E_{|i\rangle}}+\langle f |H_{4}|
f\rangle\right) \ee
where $|i\rangle$ is an intermediate state and summation is done
in all admissible channels. The first term in \eqref{eq1} gives
rise to extra logarithmic divergences. However the total answer
must be finite. This result can be achieved by imposing a unique
normal ordering prescription as  in~\cite{Astolfi:2011ju}, the
ordering prescription being the Weyl prescription.

For readers' convenience, let us briefly describe here the main
characters of this work, referring to\mycite{Astolfi:2011ju} for
exact definitions. There are four light bosonic oscillators
$a^1,a^2,\tilde{a}^1,\tilde{a}^2$; four heavy bosonic oscillators
$\hat{a}^i$, $i=1\dots 4$; light fermions $d_\alpha$, heavy
fermions $b_\alpha$ where $\alpha$ is the Dirac ten-dimensional
index. The dispersion laws are summarized in the Table \ref{dis}.
\begin{table}[h!]\caption{\label{dis}Dispersion laws}
\begin{equation*}
\begin{array}{|l|l|}\hline \hline
\mbox{State}&\mbox{Energy}\\ \hline \hline a^1,a^2&\omega_n-c/2\\
\hline \tilde{a}^1,\tilde{a}^2&\omega_n+c/2\\ \hline
\hat{a}^i&\Omega_n\\ \hline d&\omega_n\\ \hline b,
\mbox{even}&\Omega_n-c/2\\ \hline b, \mbox{odd}&\Omega_n+c/2\\
\hline\hline\end{array}
\end{equation*}
\end{table}

\noindent where the frequencies are
\beq\begin{array}{l} \omega_n=\sqrt{n^2+\frac{c^2}{4}},\\
\Omega_n=\sqrt{n^2+c^2}.
\end{array}\eeq
Referring to the ``$SU(2)\times SU(2)$ sector'' we mean states
solely consisting of either $a^1,a^2$ or
$\tilde{a}^1,\tilde{a}^2$.
Parameters of the theory are $c$, which is meant to be large
\beq c={4J\over R^2}, \eeq
the curvature radius $R$
\beq {R^2}= {4\pi\sqrt{2\lambda}}= {4\pi J \sqrt{2\lambda'}},\eeq
and the Frolov-Tseytlin coupling constant \beq
\lambda'={\lambda\over J^2}.\eeq
%
%

The main result of this Paper is in fact the extension of the results
of~\cite{Astolfi:2011ju} to the whole set of degenerate
two-oscillator light bosonic states, those states whose energies
can be also compared directly to the corresponding solutions of
the Bethe equations. These are:  8 states built up by two bosonic
oscillators and 16 made of two fermionic excitations. They have
degenerate plane-wave energy, thus the procedure for computing the
spectrum is straightforward, yet technically much more complicated
than that in~\cite{Astolfi:2011ju}: one must, as in standard
quantum mechanical perturbation theory, diagonalize the mixing
matrix of the perturbation, solve the secular equation and find
eigenvectors and eigenvalues, which are the finite size
corrections. The spectrum of such excitations, which we do not
display here for brevity (see the Tables on page 15), can be
computed exactly in $\lambda'$ and then compared with the
corresponding solutions of the asymptotic Bethe equations: this
involves the analysis of configurations carrying auxiliary roots
and thus provides a test of the Bethe program even more stringent
than the one carried out in~\cite{Astolfi:2008ji}, where the only
activated roots where the fundamental ones, carrying the physical
momentum. The Bethe equations must be solved perturbatively, by a
judicious Ansatz for the expansion of the momentum in powers of
$\lambda'={\lambda\over J^2}$ and $J$ around the asymptotic free
solution, and employing a consistent regularization technique for
the configurations with auxiliary rapidities 0 or
$\infty$~\cite{Volin:2010xz}. Having solved the Bethe equations
for the momenta, one plugs the solution in the dispersion relation
and gets the spectrum: we obtained it up to
$\mathcal{O}\left(\lambda'^4\right)$ but it can be improved with
some more computational effort. Actually we consider the
$\mathcal{O}\left(\lambda'^4\right)$ sufficient, since the
dressing phase factor interpolating from weak to strong coupling
starts at order $\lambda'^3$ and there is no physical mechanism
entering at higher orders other than those already encountered.
Thus we consider such a matching a very satisfying test to
consider it an all order result.

We defer to the main body of the Paper the detailed discussion of
the basis that diagonalizes the string theory perturbation
Hamiltonian and the corresponding Bethe Ansatz configurations. To
summarize we display here the Tables \ref{bbstba1},\ref{ffstba1}
of spectrum identifications. They refer to the Bethe
configurations in the language of~\cite{Gromov:2008qe}. The
integers $K_i$ are multiplicities of the $i$-th Bethe root. The
energies of the identified submultiplets are identical at the
order $1/J$; this is our main result, which is derived in the main
body of the paper.
\begin{table}[h!]\caption{\label{bbstba1}Boson-boson state identification}
\beq
\begin{array}{|c|ccccl|l|}\hline
\mbox{Multiplicity}&\multicolumn{5}{|c|}{\mbox{Corresponding BA
states}}&\mbox{Corresponding ST states}\\ \hline
&K_4&K_{\bar{4}}&K_3&K_2&K_1&\mbox{State nr.}\\
\hline\hline 2& 2&0&1&1&1_{\mbox{branch 1}}& 5,7\\
\hline\hline
4& 2&0&1&1&1_{\mbox{branch 2}} &2,3,6,8\\
&1&1&1&1&1_{\mbox{branch 1}}& \\
\hline\hline 2& 1&1&1&1&1_{\mbox{branch 2}}& 1,4\\
\hline\hline
\end{array}
\eeq
\end{table}
\begin{table}[h!]\caption{\label{ffstba1}
Fermion-fermion spectrum comparison}
\beq
\begin{array}{|c|ccccl|l|}\hline
\mbox{Multiplicity}& \multicolumn{5}{|c|}{\mbox{Corresponding BA
states}}&\mbox{Corresponding ST states}\\ \hline
&K_4&K_{\bar{4}}&K_3&K_2&K_1&\mbox{State nr.}\\
\hline\hline
 2& 2&0&2&2&0&23,24\\ \hline\hline
8& 1&1&2&2&0& 9, 10, 17, 18, 19, 20, 21, 22\\
& 2&0&2&1&0_{\mbox{branch 1}}&\\
& 2&0&2&1&0_{\mbox{branch 2}}&\\
& 2&0&2&0&0&\\ \hline\hline
6&1&1&2&1&0_{\mbox{branch 1}}& 11, 12, 13, 14, 15, 16\\
& 1&1&2&1&0_{\mbox{branch 2}}&\\
& 1&1&2&0&0&\\
\hline\hline
\end{array}
\eeq
\end{table}

Yet this is not the end of the story about the spectrum of the
two-oscillator bosonic states: Each of the 8 bi-bosonic and 16
bi-fermionic state has a further contribution to the energy which
is given by the same infinite sum appearing in the eqs. 1.2 and
1.3 of~\cite{Astolfi:2011ju}. The computation of such term was one
of the results of~\cite{Astolfi:2011ju} and it is therefore
appropriate to recapitulate its interpretation and inquire whether
the further developments carried out in this Paper might shed more
light about it.

There are no divergences in the eqs. 1.2 and 1.3
of~\cite{Astolfi:2011ju} due to a nontrivial, yet natural,
ordering prescription for the quantum operator associated to the
classical quartic Hamiltonian, which gives infinite sums in the
spectrum cancelling those divergencies arising from the cubic
Hamiltonian evaluated at second order in perturbation theory. This
scheme applies unchanged for the 24 bosonic states considered in
this Paper, thus providing further evidence of the naturalness of
such ordering prescription.

In~\cite{Astolfi:2011ju} it was shown that the infinite sum of
Eqs. 1.2 and 1.3  appears diagonally in the mode numbers for the
states having an arbitrary number of light bosonic oscillators.
Furthermore, if one considers a single-impurity light bosonic
state, without the level matching condition which would otherwise
have forced its mode number to be vanishing, the energy of this
state displays the same kind of contribution. Its natural
interpretation is therefore as a correction to the dispersion law
of a single magnon. This exponential one-loop effect must be
similar to the L\"uscher terms coming form a field theory or Bethe
Ansatz calculation. One should be able to directly compute it from
the L\"uscher formula (see the review~\cite{Janik:2010kd} and
references in it). This effect is yet another example of the
exponentially small finite size corrections to the magnon
dispersion relation that, for type IIA superstring on $\ads_4
\times \C P^3$, were first computed in the giant magnon limit in
\cite{Grignani:2008te} (see also
\cite{Shenderovich:2008bs,Ahn:2008hj,Abbott:2008qd,Abbott:2009um})
and derived from L\"uscher's corrections in
\cite{Bombardelli:2008qd,Lukowski:2008eq,Ahn:2008wd,Abbott:2010yb,Ahn:2010eg,Abbott:2011tp}.
Finite-size effects were also calculated for spiky strings in
$AdS_4\times \C P^3$ and for giant magnons in the presence of an
arbitrary two-form $B$ field. Alternative methods for dealing with
giant magnons on $AdS_4\times \C P^3$ by employing the so-called
dressing method were suggested in
\cite{Papathanasiou:2009en,Suzuki:2009sc,Kalousios:2009mp,Hatsuda:2009pc}.

On these grounds  it is quite crucial to inquire whether a state
built by a light non level matched fermionic oscillator indeed
displays the same kind of contribution to the spectrum: such a
computation for a fermionic  mode actually also involves the issue
of quantum ordering of classical terms quartic in the fermions,
which was not addressed in~\cite{Astolfi:2011ju} since there, they
were not relevant. The generalization to such terms of the Weyl
ordering is remarkably the unique choice leading to a finite
spectrum. This confirms the naturalness of our ordering
prescription. Even more remarkably, for each light fermionic
oscillator of a state having an arbitrary (including just one)
number of them, one obtains a contribution which is the same
infinite sum of Eqs. 1.2 and 1.3 of~\cite{Astolfi:2011ju}. This
clearly reinforces its interpretation as a finite size correction
to the magnon dispersion relation.

The light-magnon dispersion relation is fixed by symmetries of the
theory
\be \label{magnon_disp_rel} E= \sqrt{{1\over 4} +4 h^2(\lambda)
\sin^2 {p\over 2}} \ee
but the scaling function  $h(\lambda)$~\cite{Gaiotto:2008cg,
Grignani:2008is, Nishioka:2008gz} that interpolates from the
strong to the weak coupling is not. The magnon dispersion relation
\mycite{Beisert:2004hm,Beisert:2005tm} in the $\ads_5/\cft_4$
duality is
\beq E=\sqrt{1 +f(\lambda) \sin^2 {p\over 2}},\eeq
where $f(\lambda)$ happens to be equal to $\frac{\lambda}{\pi^2}$
at both  strong and weak coupling. For the $\ads_4/\cft_3$ duality
the function $h(\lambda)$ looks like $\lambda +\mathcal
O(\lambda^4)$ at weak coupling~\cite{Minahan:2008hf,
Gaiotto:2008cg, Nishioka:2008gz} and like $\sqrt{{\lambda\over
2}}+\mathcal O(\lambda^0)$ at strong
coupling~\cite{Gaiotto:2008cg, Grignani:2008te, Nishioka:2008gz}.
It has been computed up to 4 loops on the field theory side
in~\cite{Minahan:2009aq, Minahan:2009wg, Leoni:2010tb}.
Quasiclassical calculations for the spinning and folded strings
have yielded~\cite{McLoughlin:2008ms,
Alday:2008ut,Krishnan:2008zs,McLoughlin:2008he,Abbott:2010yb}
\be h(\lambda) = \sqrt{{\lambda\over 2}} +a^{\rm WS}_1 +\mathcal O
\left({1\over \sqrt{\lambda}}\right) \qquad \text{where} ~~~
a^{\rm WS}_1=- {\log{2} \over 2\pi} \,, \qquad \lambda\gg1 \ee
the superscript $\rm WS$ standing for the world-sheet.

Gromov and Vieira, on the other hand by means of the semiclassical
Bethe Ansatz ~\cite{Gromov:2008qe, Gromov:2008fy}, extrapolating
to the strong coupling of the all loop Ansatz
of~\cite{Leoni:2010tb}, obtained
 \be
h(\lambda) = \sqrt{{\lambda\over 2}}  +a^{\rm AC}_1 +\mathcal O
\left({1\over \sqrt{\lambda}}\right) \qquad \text{where} ~~~
a^{\rm AC}_1=0\,, \qquad \lambda\gg1
 \ee
 where the superscript $\rm AC$ means the algebraic curve.

The different values for $h(\lambda)$ come from different
regularizations used. If one treats all modes in a uniform way,
one gets $a_1\neq 0$. If one takes care of heavy and light modes
differently, and remembers that heavy modes are kind-of bound
states~\cite{Zarembo:2009au} of the light modes and therefore must
be cut off at a twice higher value of the momentum as the light
ones, one gets the zero $a_1$. In this work we argue that there is
a definitive evidence from the unitarity preservation requirement
to choose a unique regularization, the one with different cutoffs.

Namely, since a ``heavy-light-light'' vertex is present in the
S-matrix, the same cutoff on  mode numbers of light and heavy
states will render the regularized S-matrix non-unitary. Only
cutting the self-energy summation off in such way that preserves
unitarity at each large but finite value of the cutoff  is
acceptable. For conventional global symmetries we know that a
regularization breaking a symmetry of a theory results in an
anomaly. Unitarity is a different kind of symmetry, realized on
quantum level solely, and not at the level of the classical
Lagrangian. However, there is a great degree of resemblance
between the $\log 2$ pieces in the self-energy sums due to broken
unitarity by regularization and the presence of the anomalous
non-zero parts in otherwise classically zero divergences of
Noether currents at quantum level.

Actually the curvature corrections to the string state energies
that have been computed in~\cite{Astolfi:2011ju} and in this Paper
would notice the presence of an $a_1$ term in $h(\lambda)$. In the
BMN limit the momentum is $p=\frac{2\pi n}{J}$ and expanding for
large $J$ at nonzero $a_1$ in $h(\lambda)$ yields
\begin{eqnarray}\label{dispersioncorrected}
E&=& \sqrt{{1\over 4} +
4\left(\sqrt{\frac{\lambda}{2}}+a_1\right)^2\sin^2 {p\over
2}}\cr&\simeq&\sqrt{{1\over 4} + 2\lambda' n^2\pi^2}+
\frac{\sqrt{2}a_1}{J} \left[4 \pi ^2  n^2 \sqrt{\lambda'}-16 \pi
^4 n^4 {\lambda'}^{3/2}+96 \pi ^6  n^6
{\lambda'}^{5/2}+\mathcal{O}\left({\lambda'}^{7/2}\right)\right]\cr&&
\end{eqnarray}
namely there will be  $\frac{1}{J}= \frac{4}{c R^2}$ term with
semi-integer powers of $\lambda'$. Such a term could arise in the
finite size energies of two light magnons (see eqs. 1.2 and 1.3
of~\cite{Astolfi:2011ju}) however, with the regularization  we
use, it does not. The $\lambda'$ power expansion of the first
terms in the eqs. 1.2 and 1.3 of~\cite{Astolfi:2011ju} yields
integer powers of $\lambda'$, which are basically due to the
interactions between magnons, while the Bessel function sum with
produces non analytic terms which are exponentially suppressed
like $\sim e^{-{J\over \sqrt{\lambda}}}$, thus they are compatible
with $a_1=0$ solely.

The Paper is organized as follows.  In Section\myref{sec:spoint}
we discuss the dispersion relations of single oscillator states
with lifted level matching condition, the regularization procedure
and the unitarity argument that proves its consistency. In
Section~\ref{sec:stringcomp} we explicitly compute the energies of
the string states and construct the mixing matrix for
two-oscillator light bosonic states. In Section~\ref{sec:Bethe} we
derive the solutions of the Bethe equations corresponding to the
string states discussed in Section~\ref{sec:stringcomp}. In
Section~\ref{sec:conclusions} we draw our conclusions.


\section{Near
plane-wave limit of strings in $\ads_4\times \C P^3$:  the procedure }
\label{sec:spoint}

The type IIA  superstring of interest to us lives in the
$\ads_4\times \C P^3$ background where a two-form and four-form
Ramond-Ramond fluxes are present. The corresponding geometry is
described in the appendix (A) of\mycite{Astolfi:2011ju}. We
compute the corrections at the order $1/R^2=1/(4\pi J \sqrt{2
\lambda'})$ to the energies of the one- and two particle sectors.
We keep $J$ large, $\lambda$ large and
$\lambda'=\frac{\lambda}{J^2}=\rm{fixed}$. Thus we can say we are
in the near-BMN limit. Our corrections are perturbative quantum
mechanical corrections. Unlike the pure BMN case, we already have
some interaction in the system. Unlike the folded or rotating
string case, semiclassics are not applicable here, so exact
quantum-mechanical analysis is due. Unlike the giant magnon case,
finite size effects do not decouple from the quantum effects. Thus
our limit is in some way unique since it resides at strong
coupling, yet is perturbatively treatable.

Our basic string configuration is a point-like IIA string moving
in the $\grSU(2)\times\grSU(2)$ subsector of $\C P^3$ and along
the time direction $\R_t$ on $\ads_4$~\cite{Nishioka:2008gz,
Gaiotto:2008cg, Grignani:2008is}. The specific plane-wave
background, has been obtained in \cite{Grignani:2008is} and
discussed extensively in~\cite{Astolfi:2008ji,Astolfi:2009qh},
therefore we describe it only in the appendix (B)
of\mycite{Astolfi:2011ju}. The quantization procedure for the free
plane-wave Hamiltonian that has been done in the Section 2.1.
of\mycite{Astolfi:2011ju}. The derivation of the interaction
Hamiltonian is found in the Appendix (D)
of\mycite{Astolfi:2011ju}. All geometry, Gamma matrices, and
quantization notations are similar to those of
\mycite{Astolfi:2011ju}; with the exception of the $H_{4F}$, all
other Hamiltonian pieces are taken directly from there.

%



\subsection{Finite size dispersion relation of single
oscillator states and unitarity-preserving Regularization}
\subsection*{Single $u_4$ impurity}
We start with heavy bosonic states. Consider the single impurity
state, non level matched
\begin{equation} \label{stateu4}
|u_4\rangle = \left(a^{u_4}_n\right)^\dagger | 0 \rangle
\end{equation}

\noindent Cancellation of divergencies for bosons is a crucial
test for the validity of our theory. To illustrate how
divergencies cancel in the dispersion relation for the fourth
heavy boson, we show the partial contributions of different
sectors in the Table\myref{boscanc} below. The boson energy is
\beq E_n^{u_4}=\Omega_n +\frac{1}{R^2\Omega_n}\sum_q
\epsilon_q^{u_4}.\eeq
\begin{table}[h!]
\caption{\label{boscanc}Cancellation  of divergencies for bosons}
\begin{equation*}\displaystyle
\begin{array}{|l|l|}\hline
\hline \multicolumn{1}{|c|}{\mbox{Hamiltonian
piece}}&\multicolumn{1}{|c|}{\epsilon_q^{u_4}}
\\[2pt]\hline
H_3^2& -\frac{4 q^2}{c \omega _q}+\frac{n^2}{\omega _q \omega
_{n+q}}+\frac{2 n
   q}{\omega _q \omega _{n+q}} \\[2pt]\hline
H_{\hat{u}_4\hat{u}_4BB}^{light}& -\frac{8 n^2 q^2}{c^3 \omega
_q}-\frac{8 n q \Omega _n}{c^3}-\frac{2
   n^2}{c \omega _q}-\frac{4 q^2}{c \omega _q} \\[2pt]\hline
H_{\hat{u}_4\hat{u}_4BB}^{heavy}& \hphantom{-}\frac{6 n q \Omega
_n}{c^3}
-\frac{6 q^2 \Omega _n^2}{c^3 \Omega _q} \\[2pt]\hline
H_{4\hat{u}_4}& -\frac{2 n^2 q^2}{c^3 \Omega _q}+\frac{2 n q
\Omega _n}{c^3}-\frac{2
   n^2}{c \Omega _q}-\frac{2 q^2}{c \Omega _q} \\[2pt]\hline
H_{2B2F}^{light}& \hphantom{-}\frac{8 q^2 \Omega _n^2}{c^3 \omega
_q}-\frac{8 n q  \Omega _n}{c^2}
\\[2pt]\hline
H_{2B2F}^{heavy}& \hphantom{-}\frac{8q^2 \Omega _n^2}{c^3 \Omega
_q}-\frac{8 n q \Omega_n}{c^2}+\frac{4 n^2}{c\Omega _q}\\[2pt]
\hline\hline
\mbox{Total}&\hphantom{-}\frac{2n^2}{c}\left(\frac{1}{\Omega_q}
-\frac{1}{\omega_q}\right)
\\[2pt] \hline\hline
\end{array}
\end{equation*}
\end{table}

\noindent The ``total'' line of the table refers to the sum both
over partial contributions and the summation mode index. This lets
the expressions be additionally simplified, since the dumb
variable allows constant shifts, leading to extra cancellations.
Also note exact cancellation of quadratic divergencies. By summing
over partial channels shown above the dispersion relation up to
finite size becomes
\begin{eqnarray}  \label{disprelu4}
& & E^{u_4}= \sqrt{1+\frac{n^2}{c^2}}+\frac{2 n^2}{c \Omega_n
R^2}\left(\sum_{q=-2 N}^{2 N}\frac{1}{\Omega_q}-\sum_{q=-N}^N
\frac{1}{\omega_q}\right)+\cr && \frac{1}{c R^2
\Omega_n}\sum_{q=-N}^N\bigg\{4 \frac{q^2}{\omega_q}-
\left(2\omega_q+\omega_{q+n}
+\omega_{q-n}\right)+\frac{c^2}{4}\left(\frac{2}{\omega_q}
+\frac{1}{\omega_{q+n}}+\frac{1}{\omega_{q-n}}\right)\cr &&-
\frac{c}{2}\left(\frac{\omega_{n+q}}{\omega_q}-
\frac{\omega_q}{\omega_{n+q}}+\frac{\omega_{-n+q}}{\omega_q}-
\frac{\omega_q}{\omega_{-n+q}}\right)\bigg\}
\end{eqnarray}

\noindent We already know very well how to treat the sum in the
first line in eq.\eqref{disprelu4} since it is exactly the one
appearing in\mycite{Astolfi:2011ju}, giving rise to the Bessel
functions series. For this state, this sum appears uniquely from
contributions due to the quartic Hamiltonian. The other terms in
the equation are organized as follows: their cutoff is $N$ because
the sum is over a light mode.

Since the sum in the second and third lines of  \eqref{disprelu4}
are convergent and have the same cutoff $N$, we can safely send it
to infinity and performing some shifts we can see that all the
terms sum up to zero. Therefore the dispersion relation is rather
the following simpler one:
\begin{equation} \label{disprelu4bis}
E^{u_4}=\sqrt{1+\frac{n^2}{c^2}}+\frac{2 n^2}{c  \Omega_n
R^2}\left(\sum_{q=-2 N}^{2 N}\frac{1}{\Omega_q}-\sum_{q=-N}^N
\frac{1}{\omega_q}\right)
\end{equation}

\subsection*{Single $u_1$ impurity}

The interaction Hamiltonian is not explicitly invariant with
regard to $u_4\to u_i$ replacement, since the fourth direction is
special it belongs to $\C P^3$ while $u_i\in \ads_4$ for
$i=1,2,3$. Therefore we must consider now the single impurity
heavy state with a $u_1$ oscillator separately
\begin{equation} \label{stateu1}
| u_1 \rangle = \left(a^{u_1}_n\right)^\dagger | 0 \rangle
\end{equation}

\noindent We see by an explicit calculation that its dispersion
relation up to finite size is the same as for the $u_4$ single
oscillator state:
\begin{equation} \label{disprelu1}
E^{u_1}=\sqrt{1+\frac{n^2}{c^2}} +\frac{2 n^2}{c \Omega_n
R^2}\left(\sum_{q=-2 N}^{2 N}\frac{1}{\Omega_q}-\sum_{q=-N}^N
\frac{1}{\omega_q}\right)
\end{equation}

\noindent By virtue of the same argument as above we can see that
all divergencies cancel, whereas the remaining term contains only
an exponentially small correction in $J$.

\subsection*{Fermions}
Consider now the fermionic states: the light one

\beq |d\rangle= d_{\alpha n}^\dagger|0\rangle,\eeq

\noindent and the heavy one \beq |b\rangle=b_{\alpha
n}^\dagger|0\rangle. \eeq

\noindent We check the fermion dispersion relation perturbatively
and demonstrate the results in the Table\myref{canc}.
\begin{table}[h!]\caption{\label{canc}Cancellation of divergencies
for fermions. Separate sectors give divergent results, the remnant
is finite.}
\begin{equation*} \displaystyle
\begin{array}{|l|l|l|}\hline\hline
\multicolumn{1}{|c|}{\mbox{Hamiltonian
piece}}&\multicolumn{1}{|c|}{\mbox{Light state energy correction
$\epsilon^{d}$ }}&\multicolumn{1}{|c|}{\mbox{Heavy state state
energy correction $\epsilon^{b}$}}
\\[2pt] \hline \displaystyle H^2_{3\,\,\mbox{light loop}}
& -\frac{3 n^2}{16 c \omega _q}-\frac{3 q^2}{16 c \omega
_q}+\frac{3 q^2}{16
   c \Omega _q} & -\frac{n^2}{8 c \omega _q} \\[2pt] \hline
\displaystyle H^2_{3\,\,\mbox{heavy loop}}& -\frac{n^2}{4 c
\Omega_q} +\frac{q^2}{16 c \omega _q}-\frac{9 c}{64 \omega_q}
-\frac{q^2}{16 c \Omega _q}-\frac{9 c}{32 \Omega_q}
 & \hphantom{-}0 \\[2pt] \hline
 \displaystyle
H_{2B2F\,\,\mbox{light loop}}& -\frac{2 q^2 n^2}{c^3 \omega
_q}-\frac{n^2}{2 c \omega _q}-\frac{q^2}{16 c
   \omega _q}+\frac{9 c}{64 \omega _q} & -\frac{4 q^2 n^2}{c^3 \omega
   _q}-\frac{n^2}{c \omega _q}-\frac{5 q^2}{2 c \omega _q}-\frac{9 c}{16
   \omega _q} \\[2pt] \hline
 \displaystyle
H_{2B2F\,\,\mbox{heavy loop}}& -\frac{2 q^2 n^2}{c^3 \Omega
_q}-\frac{n^2}{2 c \Omega _q}-\frac{3 q^2}{16
   c \Omega _q} & -\frac{4 q^2 n^2}{c^3 \Omega _q}-\frac{n^2}{c \Omega
   _q}-\frac{2 q^2}{c \Omega _q} \\[2pt] \hline
 \displaystyle
H_{4F\,\,\mbox{light loop}}& \hphantom{-}\frac{2 q^2 n^2}{c^3
\omega _q}+\frac{5 n^2}{16 c \omega _q}+\frac{5
   q^2}{16 c \omega _q} & \hphantom{-}\frac{4 q^2 n^2}{c^3 \omega _q}
+\frac{n^2}{8 c
   \omega _q}+\frac{5 q^2}{2 c \omega _q}+\frac{9 c}{16 \omega _q}
   \\[2pt] \hline
 \displaystyle
H_{4F\,\,\mbox{heavy loop}}&
 -\frac{n^2}{8 c \omega _q}+\frac{2 q^2 n^2}{c^3 \Omega _q}+\frac{5 n^2}{4 c
   \Omega _q}-\frac{q^2}{8 c \omega _q}+\frac{q^2}{16 c \Omega _q}+\frac{9
   c}{32 \Omega _q} & \hphantom{-}\frac{4 q^2 n^2}{c^3 \Omega _q}
   +\frac{2 n^2}{c \Omega
   _q}+\frac{2 q^2}{c \Omega _q} \\[2pt] \hline\hline
    \displaystyle
   \mbox{Total}&
 \hphantom{-}\frac{n^2}{2 c \Omega _q}-\frac{n^2}{2 c \omega _q} &
 \hphantom{-}\frac{n^2}{c \Omega
   _q}-\frac{n^2}{c \omega _q}\\[2pt]  \hline\hline
\end{array}
\end{equation*}\end{table}

\noindent The superficial divergences present in the loop
contributions to fermions do cancel indeed, only a finite
exponentially suppressed part (as $e^{-const J}$) remaining. Here
we demonstrate how various contributions cancel in order to leave
a finite piece only. For light states
\beq E^d=\omega_n +\frac{1}{R^2\omega_n}\sum_q \epsilon_q^{d},\eeq
for heavy states
\beq E^b=\Omega_n +\frac{1}{R^2\Omega_n}\sum_q \epsilon_q^{b}.\eeq
The partial (taken over separate channels) $\epsilon^{d}_q$ and
$\epsilon^{b}_q$ are given in the Table\myref{canc}.

\subsection*{Unitarity-preserving Regularization}
{ The sums of the light and heavy self-energies $\sum_q
\epsilon_q^{d},\,\, \sum_q \epsilon_q^{b}$ are convergent, but
have to be regularized to be ascribed a numerical value. We use
here the most natural ``algebraic-curve'' regularization
prescription suggested by the form of the cubic
Hamiltonian~\cite{Astolfi:2011ju}. That is, we cut the heavy modes
at a cutoff $2N$, and the light modes at a cutoff $N$, where $N$
is afterwards sent to infinity. The arguments that have been used
in literature for this regularization have been reiterated below
and will be discussed yet once more in the Conclusion; here we
wish to bring in a very generic argument, which leaves this
``unequal-frequency'' regularization as the only permissible one.
Zarembo has shown\mycite{Zarembo:2009au} that the heavy-to-light
vertex is organized in our theory in such a way that the light
two-particle cut starts exactly in the point where the heavy
particle pole is. Thus an on-shell decay heavy-into-two-light
modes is seen to be possible. Consider now the most general
requirement of validity of a quantum field theory, the equation on
the unitarity of the S-matrix
\beq SS^\dagger=1\eeq
Let us write it down more explicitly in the 1-particle heavy
sector (indices $1,i$) and the 2-particle sector (indices $2,jk$)
\beq S_{im}^{1,1}S_{mi\prime}^{1,1\dagger}
+S_{i,jk}^{1,2}S_{jk,i\prime}^{1,2\dagger}=1_{i,i\prime}^{1,1}
\eeq
where summation is meant over the repeating Hilbert space indices.
The second part of the left-hand side must be taken into account
due to the mentioned result by Zarembo. The relation is valid at
any mode number. Suppose we regularize the theory now. This
effectively means that both for the unit operator in the Hilbert
space and for the S-matrices all elements above some $N_{cutoff}$
are filled in with zeros. Suppose that $N_{cutoff}$ is special for
each of the sectors. Thus we have below the cutoffs

\beq \left.
S_{2n,2n}^{1,1}S_{2n,2n}^{1,1\dagger}\right|_{2n<N_{cutoff}^{heavy}}
+\left.S_{2n,n\,n}^{1,2}S_{n\,n,2n}^{1,2\dagger}\right|_{n<N_{cutoff}^{light}}=
\left.1_{2n,2n}^{1,1}\right|_{2n<N_{cutoff}^{heavy}}, \eeq

\beq \left.
S_{2n,2n}^{1,1}S_{2n,2n}^{1,1\dagger}\right|_{2n>N_{cutoff}^{heavy}}
+\left.S_{2n,n\,n}^{1,2}S_{n\,n,2n}^{1,2\dagger}\right|_{n>N_{cutoff}^{light}}=
0. \eeq

Then we  can see that the only way to comply with unitarity is to
impose $N_{cutoff}^{heavy}=2 N_{cutoff}^{light}$. The intuitive
way to understand this physics is very simple: if you cut the
Hilbert space off at an arbitrary energy, the heavy modes will
decay into nothing. This is precisely what we usually understand
as a non-unitary theory - a non-unity-normalized total probability
of an inclusive process (in our case, it's ``heavy mode into
something'').}

Thus we impose  the regularization and calculate the sums above as
was done in\mycite{Astolfi:2011ju}. We obtain \beq\delta E \sim
e^{-J/\sqrt{2\lambda}},\eeq which means there are no power
corrections to the energy.   From these results we get in the
strong-coupling limit the function $h(\lambda)$, parameterized as
\beq h(\lambda)=\sqrt{\frac{\lambda}{2}}+a_1+...\, ,\eeq that \beq
a_1=0, \eeq supporting the result from the Bethe Ansatz algebraic
curve.


\section{Finite-size mixing matrix for two-oscillator light bosonic states}
\label{sec:stringcomp}

The mixing matrix between two-oscillator  bosonic states having
degenerate plane-wave energies is \be \label{mixing}
M_{\rm{mix}}^{ij}=\sum_{|i\rangle}\frac{\langle e_i | H_{(3)} | i
\rangle \langle i H_{(3)} | e_j \rangle }{E^{(0)}_{|e_i
\rangle}-E^{(0)}_{|i\rangle}}+\langle e_i |H_{4}| e_j\rangle \ee

\noindent Solving the secular equation for $M_{\rm{mix}}^{ij}$
one gets the the eigenvectors and the eigenvalues, i.e. the finite
size corrections to the spectrum. The four single-oscillator light
bosonic states are \beq\{ a_{n}^{1\dagger}|0\rangle,
\tilde{a}_{n}^{1\dagger}|0\rangle, a_{n}^{2 \dagger}|0\rangle,
\tilde{a}_{n}^{2\dagger}|0\rangle \},\eeq and the
single-oscillator light fermionic states are $d_{\alpha,
n}^\dagger|0\rangle$.

\noindent We can build eight degenerate states with two bosonic
oscillators and sixteen physical states with two fermionic
oscillators, all of them having plane-wave energy $2 \omega_n$, in
units of $c$. There are other eight bibosonic states, which are
non-degenerate and have been considered in\mycite{Astolfi:2011ju}.
An educated guess, on the grounds of the symmetries of the
Hamiltonian, on the choice of the basis $| e_i \rangle$ shall
sharpen the computation considerably and that's what we are up to.
One could naively build these four bosonic states:
\beq\begin{array}{l}\label{basis3}
v^1_n=a_{n}^{1\dagger} \tilde{a}_{-n}^{1\dagger}|0\rangle,\\
v^2_n=a_{n}^{1\dagger}\tilde{a}_{-n}^{2\dagger}|0\rangle,\\
v^3_n=a_{n}^{2\dagger}\tilde{a}_{-n}^{1\dagger}|0\rangle,\\
v^4_n=a_{n}^{2\dagger}\tilde{a}_{-n}^{2\dagger}|0\rangle.
\end{array}
\eeq where $n>0$, and equally states with $n\to -n$. The true
basis should however possess definite parities with respect to
$\mathbb{Z}_2$ symmetries: the momentum reflection symmetry $P_n:
n\to -n$, the symmetry between the two $SU(2)$'s $P_a: a^1\to
a^2$, and the symmetry $\tilde{P}: a_i\to \tilde{a}_i$. Such
states are easily constructed as follows: first symmetrize and
antisymmetrize in $P_a$

\beq\label{basis2}
\begin{array}{l}
s_n=\frac{1}{\sqrt{2}}(v^1_n+v^4_n)\\
p_n=\frac{1}{\sqrt{2}}(-v^1_n+v^4_n)\\
q_n=\frac{1}{\sqrt{2}}(v^2_n+v^3_n)\\
r_n=\frac{1}{\sqrt{2}}(-v^2_n+v^3_n).
\end{array}
\eeq

\noindent The full basis of Lorenzian spin zero light
two-oscillator boson-boson tree-level degenerate states has then
dimension eight and is, after
 decomposition into $P_n$ even and odd states:

\beq\label{basis1}\begin{array}{|l|l|r|r|r|}\hline
\mbox{state}&\mbox{definition}&P_n&\tilde{P}&P_a\\ \hline
u_1&\frac{1}{\sqrt{2}}(s_n+s_{-n})&1&1&1\\ \hline
u_2&\frac{1}{\sqrt{2}}(-s_n+s_{-n})&-1&-1&1\\ \hline
u_3&\frac{1}{\sqrt{2}}(p_n+p_{-n})&1&1&-1\\ \hline
u_4&\frac{1}{\sqrt{2}}(-p_n+p_{-n})&-1&-1&-1\\ \hline
u_5&\frac{1}{\sqrt{2}}(q_n+q_{-n})&1&1&1\\ \hline
u_6&\frac{1}{\sqrt{2}}(-q_n+q_{-n})&-1&-1&1\\ \hline
u_7&\frac{1}{\sqrt{2}}(r_n+r_{-n})&1&-1&-1\\ \hline
u_{8}&\frac{1}{\sqrt{2}}(-r_n+r_{-n})&-1&1&-1\\ \hline
\end{array}\eeq

\noindent The two-fermion-oscillator states are of the type
$d^\dagger_\alpha A_{\alpha\beta} d^\dagger_\beta |0\rangle $, where the ones out of them with
zero AdS spin $s$ can potentially mix with the light bosonic
states as well. $A_{\alpha\beta}$ is an arbitrary matrix
with fermionic indices. Choosing the linearly independent states
by the following projection criteria:

\beq \begin{array}{l}\Gamma_{11}^T A \Gamma_{11}\neq 0,\\ \\
\Gamma_{+}^T A \Gamma_{+}\neq 0,\\ \\
\mathcal{P}^T A \mathcal{P}\neq 0\\ \\
\end{array} \eeq

\noindent we find the fermionic-fermionic basis has dimension sixteen, as physically expected since the light physical degrees of freedom of each 32-dimensional spinor of our theory are four.
Therefore, the total basis in the mixing sector has 24 dimensions;
the states $u_{9..24}=\{d^+ A_{i} d^+|0\rangle\}$ can be
chosen as \beq\begin{array}{rcl}
 A_{i=9\dots 24}&=&\{\frac{1}{2}, \frac{1}{2}\Gamma_{56}, \frac{1}{2}\Gamma_{12},
 \frac{1}{2}\Gamma_{13},
 \frac{1}{2}\Gamma_{23},\\ \\ &&\frac{1}{2}\Gamma_{1256},
 \frac{1}{2}\Gamma_{1356}, \frac{1}{2}\Gamma_{2356},
     \frac{1}{2}\Gamma_{1457}, \frac{1}{2}\Gamma_{2457},\\ \\ &&
     \frac{1}{2}\Gamma_{3457}, \frac{1}{2}\Gamma_{1467},
     \frac{1}{2}\Gamma_{2467},
     \frac{1}{2}\Gamma_{3467}, \frac{1}{2}\Gamma_{0579},
     \frac{1}{2}\Gamma_{0679}\}.\end{array}\eeq
where the index $i$ numbering basis states runs from 9 to 24. It
can be explicitly seen that this basis is orthonormal.

The computation of the mixing matrix is standard. When taking the
matrix elements, we consider only the contraction combinations in
which the oscillators of the quartic Hamiltonian are all
contracted with those of the external states. For the cubic
Hamiltonian evaluated at second order in perturbation theory, we
keep only the combinations of contractions which lead to enough
delta's on the mode numbers as to determine all the free indices
of the intermediate states as functions of the external state
ones, $n$ and $-n$. We are then discarding all the contractions
bringing contributions to the spectrum as infinite sums over a
free summand: these have actually been addressed separately and
they lead to the finite-size correction to the dispersion
relation, which is a peculiarity of this theory, as discussed in
the Introduction and in~\cite{Astolfi:2011ju}.

\begin{figure}[h]\begin{center}
\includegraphics[height = 3cm, width=5cm]{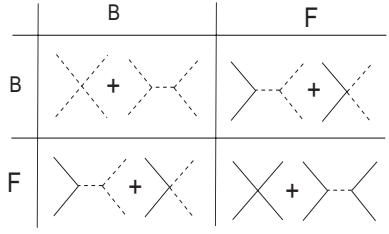}
\caption{\label{sectors}Schematic diagram of the contributing
sectors of string Hamiltonian}\end{center}
\end{figure}

\noindent Schematically, we may picture the mixing matrix as
follows
\beq\begin{array}{c|cc} &B&F\\
\hline B&H_{4B}+H_{3B}^2&H_{2B2F}+H_{FFB}H_{BBF}\\ \\
F&H_{2B2F}+H_{FFB}H_{BBF}&H_{4F}+H_{FFB}^2\\ \\
\end{array}
\eeq where $F$ and $B$ represent the two-fermion and two-boson
oscillator states. This Hamiltonian is symbolically depicted
in\myfigref{sectors}.

\noindent The corresponding pieces of the mixing matrix are shown
below.

\noindent The four-boson contribution to the BB sector is the
following $8\times 8$ matrix given in terms of the basis states
$u_1\dots u_8$: \beq H_{4B} = \frac{2 n^2}{c
\omega_n^2}\mathrm{Diagonal}\left(
\begin{array}{cccccccc}
 -\frac{4 n^2}{c^2}-4,&-\frac{4 n^2}{c^2}-3,& -\frac{4 n^2}{c^2}-4 ,&
-\frac{4 n^2}{c^2}-5 ,& -\frac{4 n^2}{c^2} ,& -\frac{4 n^2}{c^2}-1 ,
&-\frac{4 n^2}{c^2} ,& -\frac{4 n^2}{c^2}-1
\end{array}
\right). \eeq

\noindent The $H_{3B}$ contributes via three types of intermediate
states \beq H_{3B}^2=\sum_{j=1}^3 H_{3B_{j}},\eeq

\noindent where the intermediate states $s_j$ are: \beq\begin{array}{l}
|s_1\rangle=|\hat{a}^{4\dagger}_0 | 0  \rangle,\\
|s_2\rangle=|\hat{a}^{4\dagger}_{-a-b} a^\dagger_a \tilde{a}^\dagger_b | 0 \rangle,\\
|s_3\rangle=|\hat{a}^{4\dagger}_{-a-b-c-d} a^\dagger_a \tilde{a}^\dagger_b
a^\dagger_c \tilde{a}^\dagger_d | 0 \rangle.\\
 \end{array}\eeq

\noindent The following matrix (in terms of the same bosonic basis
as above) is the contribution of the first channel \beq
H_{3B_{1}}^2=\frac{4 n^2}{c \omega_n^2} \mathrm{Diagonal}\left(
\begin{array}{cccccccc}
 0,&  0, &\omega_n+\frac{c}{2}, & 0, & 0, & 0, & 0, & 0
\end{array}
\right). \eeq

\noindent The second channel gives \beq H_{3B_{2}}^2=\frac{2
n^2}{c \omega_n^2} \mathrm{Diagonal}\left(
\begin{array}{cccccccc}
 0, &  1, & 0, & 1, & 0, & -1, & 0,  &
 -1,
\end{array}
\right).\eeq

\noindent Finally the third channel gives \beq H_{3B_{3}}^2=- \frac{4
n^2}{c \omega_n^2}\mathrm{Diagonal}\left(
\begin{array}{cccccccc}
 0, & 0, &  \omega_n-\frac{c}{2}, & 0, & 0, & 0, & 0, & 0
\end{array}
\right).\eeq

Switch now to the two fermionic oscillators. The quartic purely
fermionic Hamiltonian is
\begin{equation}
\label{pH4F}
\begin{array}{l} \ds
\CH_{4,F} = - \frac{i}{24} \Big( \bar{\theta} \Gamma_{11} \Gamma_+
\CM^2 \theta' + \bar{\theta} \Gamma_+ \CM^2 \Gamma_{11} \theta'
\Big)
- \frac{1}{2c} ( A_{+,\sigma}^2 - \tilde{A}_{+,\sigma}^2 )
\\[2mm] \ds
- \frac{1}{4} A_{+,\sigma} ( \tilde{C}_{+-} + \tilde{B}_{+56} +
\tilde{B}_{+78} )
+ \frac{1}{4} \tilde{A}_{+,\sigma} ( C_{+-} - C_{++} + B_{+56} +
B_{+78} )
\\[2mm] \ds
- \frac{c}{8} \sum_{i=1}^4 C_{+i}^2
- \frac{c}{32} \sum_{i=5}^8 \Big[ 2 C_{+i} - s_i B_{+4i} +
\frac{1}{2} \sum_{j=5}^8 \epsilon_{ij} B_{+-j} \Big]^2
\end{array}.
\end{equation}
Unlike the other sectors where we have referred the reader to
\mycite{Astolfi:2011ju} for the Hamiltonian expression, we show
$\CH_{4,F}$ explicitly, since we correct a misprint  of the
earlier version.

In the basis $u_{9}\dots u_{24}$ the $16\times 16$ matrix of the
mixing in the fermionic sector is \beq H_{4F}=-\frac{2 n^2}{c
\omega_n^2}
\begin{array}{cccccccc}
\mathrm{Diagonal}\left(\frac{4 n^2}{c^2}+\frac{3}{2},\right. &
\frac{4 n^2}{c^2}-\frac{3}{2}, & \frac{4 n^2}{c^2}+\frac{5}{2}, &
\frac{4 n^2}{c^2}+\frac{5}{2}, & \frac{4 n^2}{c^2}+\frac{5}{2}, &
\frac{4 n^2}{c^2}+\frac{3}{2}, & \frac{4 n^2}{c^2}+\frac{3}{2}, & \frac{4 n^2}{c^2}+\frac{3}{2}, \\
\frac{4 n^2}{c^2}+\frac{1}{2}, & \frac{4 n^2}{c^2}+\frac{1}{2}, &
\frac{4 n^2}{c^2}+\frac{1}{2}, & \frac{4 n^2}{c^2}+\frac{1}{2}, &
\frac{4 n^2}{c^2}+\frac{1}{2}, & \frac{4 n^2}{c^2}+\frac{1}{2}, &
\frac{4 n^2}{c^2}-\frac{1}{2}, & \left. \frac{4
n^2}{c^2}-\frac{1}{2}\right)
\end{array}
.\eeq
The mixing term in the FF sector coming from the $H_{FFB}^2$ is
the following:
\beq  H_{FFB}^2=\frac{2 n^2}{c
\omega_n^2}\mathrm{Diagonal}\left(
\begin{array}{cccccccccccccccc}
\frac{3}{2},& -\frac{3}{2}, & \frac{1}{2}, & \frac{1}{2},&
\frac{1}{2}, & -\frac{1}{2}, & -\frac{1}{2},  & -\frac{1}{2},  &
\frac{1}{2}, & \frac{1}{2}, & \frac{1}{2}, &\frac{1}{2},&
\frac{1}{2}, &\frac{1}{2} ,& \frac{3}{2}, & \frac{3}{2}
\end{array}
\right). \eeq \noindent This term comes from the long intermediate
channel; the short channel yields identically 0.
Of the 16 fermionic states, $u_{9}\dots u_{24}$ only states $u_9$,
$u_{10}$ have Lorenzian $AdS$ spin $s=0$ and thus could have in
principle mixed with bibosonic states, all of which have spin 0.
This mixing indeed is vanishing.
There could be in principle a mixed term $H_{BBB} H_{BFF}$, with a
bosonic zero mode intermediate state as shown
in\myfigref{sectors}, but explicit calculation shows it is zero.
Summing the contributions we get the full mixing matrix, which our
judicious choice of the basis has automatically brought to a
diagonal form. We have therefore obtained the set of eigenstates
and eigenvalues exact in $\lambda'$. The following is the
finite-size spectrum of two  bosonic oscillations, where, on the
rightmost column, in order for further comparison with the Bethe
Ansatz, we have expressed the spectrum in terms of $\lambda'$ and
$J$, expanding up to fourth order in $\lambda'$: see table
\ref{tbo}.
\begin{table}[h!]\caption{Finite-size spectrum of two
  bosonic oscillations\label{tbo}}
\begin{equation*}
\begin{array}{|c|c|c|} \hline
\mbox{state}&\mbox{spectrum}&\mbox{expansion of the spectrum}\\ \hline
u_1&-\frac{8 n^2 \Omega_n^2}{c^3 R^2 \omega_n^2}&\frac{1}{J}\left( -16 n^2 \pi ^2 \lambda '+96 n^4 \pi ^4 \lambda'^2 -768 n^6 \pi^6 \lambda'^3
+6144 n^8 \pi ^8 \lambda'^4+\dots\right)\\ \hline
u_2&-\frac{4 n^2\left(c^2+2 n^2\right)}{c^3 R^2 \omega_n^2}&\frac{1}{J}\left( -8 n^2 \pi ^2 \lambda '+32 n^4 \pi ^4 \lambda'^2 -256 n^6 \pi^6 \lambda'^3
+2048 n^8 \pi ^8 \lambda'^4+\dots\right)\\ \hline
u_3&-\frac{4 n^2\left(c^2+2 n^2\right)}{c^3 R^2 \omega_n^2}&\frac{1}{J}\left( -8 n^2 \pi ^2 \lambda '+32 n^4 \pi ^4 \lambda'^2 -256 n^6 \pi^6 \lambda'^3
+2048 n^8 \pi ^8 \lambda'^4+\dots\right)\\ \hline
u_4&-\frac{8 n^2 \Omega_n^2}{c^3 R^2 \omega_n^2}&\frac{1}{J}\left( -16 n^2 \pi ^2 \lambda '+96 n^4 \pi ^4 \lambda'^2 -768 n^6 \pi^6 \lambda'^3
+6144 n^8 \pi ^8 \lambda'^4+\dots\right)\\ \hline
u_5&-\frac{8 n^4}{c^3 R^2 \omega_n^2}&\frac{1}{J}\left(-32 n^4 \pi ^4 \lambda'^2 +256 n^6 \pi^6 \lambda'^3
-2048 n^8 \pi ^8 \lambda'^4+\dots\right)\\ \hline
u_6&-\frac{4 n^2\left(c^2+2 n^2\right)}{c^3 R^2 \omega_n^2}&\frac{1}{J}\left( -8 n^2 \pi ^2 \lambda '+32 n^4 \pi ^4 \lambda'^2 -256 n^6 \pi^6 \lambda'^3
+2048 n^8 \pi ^8 \lambda'^4+\dots\right)\\ \hline
u_7&-\frac{8 n^4}{c^3 R^2 \omega_n^2}&\frac{1}{J}\left(-32 n^4 \pi ^4 \lambda'^2 +256 n^6 \pi^6 \lambda'^3
-2048 n^8 \pi ^8 \lambda'^4+\dots\right)\\ \hline
u_8&-\frac{4 n^2\left(c^2+2 n^2\right)}{c^3 R^2 \omega_n^2}&\frac{1}{J}\left( -8 n^2 \pi ^2 \lambda '+32 n^4 \pi ^4 \lambda'^2 -256 n^6 \pi^6 \lambda'^3
+2048 n^8 \pi ^8 \lambda'^4+\dots\right)\\ \hline
\end{array}
\end{equation*}
\end{table}
\noindent The finite-size spectrum of the two fermionic
oscillations is given in the Table \ref{tfo} below.
\begin{table}[h!]
\caption{Finite-size spectrum of two
  fermionic oscillations\label{tfo}}
\begin{equation*}
\begin{array}{|c|c|c|} \hline
\mbox{state}&\mbox{spectrum}&\mbox{expansion of the spectrum}\\ \hline
u_9&-\frac{8 n^4}{c^3 R^2 \omega_n^2}&\frac{1}{J}\left(-32 n^4 \pi ^4 \lambda'^2 +256 n^6 \pi^6 \lambda'^3
-2048 n^8 \pi ^8 \lambda'^4+\dots\right)\\ \hline
u_{10}&-\frac{8 n^4}{c^3 R^2 \omega_n^2}&\frac{1}{J}\left(-32 n^4 \pi ^4 \lambda'^2 +256 n^6 \pi^6 \lambda'^3
-2048 n^8 \pi ^8 \lambda'^4+\dots\right)\\ \hline
u_{11}&-\frac{4 n^2\left(c^2+2 n^2\right)}{c^3 R^2 \omega_n^2}&\frac{1}{J}\left( -8 n^2 \pi ^2 \lambda '+32 n^4 \pi ^4 \lambda'^2 -256 n^6 \pi^6 \lambda'^3
+2048 n^8 \pi ^8 \lambda'^4+\dots\right)\\ \hline
u_{12}&-\frac{4 n^2\left(c^2+2 n^2\right)}{c^3 R^2 \omega_n^2}&\frac{1}{J}\left( -8 n^2 \pi ^2 \lambda '+32 n^4 \pi ^4 \lambda'^2 -256 n^6 \pi^6 \lambda'^3
+2048 n^8 \pi ^8 \lambda'^4+\dots\right)\\ \hline
u_{13}&-\frac{4 n^2\left(c^2+2 n^2\right)}{c^3 R^2 \omega_n^2}&\frac{1}{J}\left( -8 n^2 \pi ^2 \lambda '+32 n^4 \pi ^4 \lambda'^2 -256 n^6 \pi^6 \lambda'^3
+2048 n^8 \pi ^8 \lambda'^4+\dots\right)\\ \hline
u_{14}&-\frac{4 n^2\left(c^2+2 n^2\right)}{c^3 R^2 \omega_n^2}&\frac{1}{J}\left( -8 n^2 \pi ^2 \lambda '+32 n^4 \pi ^4 \lambda'^2 -256 n^6 \pi^6 \lambda'^3
+2048 n^8 \pi ^8 \lambda'^4+\dots\right)\\ \hline
u_{15}&-\frac{4 n^2\left(c^2+2 n^2\right)}{c^3 R^2 \omega_n^2}&\frac{1}{J}\left( -8 n^2 \pi ^2 \lambda '+32 n^4 \pi ^4 \lambda'^2 -256 n^6 \pi^6 \lambda'^3
+2048 n^8 \pi ^8 \lambda'^4+\dots\right)\\ \hline
u_{16}&-\frac{4 n^2\left(c^2+2 n^2\right)}{c^3 R^2 \omega_n^2}&\frac{1}{J}\left( -8 n^2 \pi ^2 \lambda '+32 n^4 \pi ^4 \lambda'^2 -256 n^6 \pi^6 \lambda'^3
+2048 n^8 \pi ^8 \lambda'^4+\dots\right)\\ \hline
u_{17}&-\frac{8 n^4}{c^3 R^2 \omega_n^2}&\frac{1}{J}\left(-32 n^4 \pi ^4 \lambda'^2 +256 n^6 \pi^6 \lambda'^3
-2048 n^8 \pi ^8 \lambda'^4+\dots\right)\\ \hline
u_{18}&-\frac{8 n^4}{c^3 R^2 \omega_n^2}&\frac{1}{J}\left(-32 n^4 \pi ^4 \lambda'^2 +256 n^6 \pi^6 \lambda'^3
-2048 n^8 \pi ^8 \lambda'^4+\dots\right)\\ \hline
u_{19}&-\frac{8 n^4}{c^3 R^2 \omega_n^2}&\frac{1}{J}\left(-32 n^4 \pi ^4 \lambda'^2 +256 n^6 \pi^6 \lambda'^3
-2048 n^8 \pi ^8 \lambda'^4+\dots\right)\\ \hline
u_{20}&-\frac{8 n^4}{c^3 R^2 \omega_n^2}&\frac{1}{J}\left(-32 n^4 \pi ^4 \lambda'^2 +256 n^6 \pi^6 \lambda'^3
-2048 n^8 \pi ^8 \lambda'^4+\dots\right)\\ \hline
u_{21}&-\frac{8 n^4}{c^3 R^2 \omega_n^2}&\frac{1}{J}\left(-32 n^4 \pi ^4 \lambda'^2 +256 n^6 \pi^6 \lambda'^3
-2048 n^8 \pi ^8 \lambda'^4+\dots\right)\\ \hline
u_{22}&-\frac{8 n^4}{c^3 R^2 \omega_n^2}&\frac{1}{J}\left(-32 n^4 \pi ^4 \lambda'^2 +256 n^6 \pi^6 \lambda'^3
-2048 n^8 \pi ^8 \lambda'^4+\dots\right)\\ \hline
u_{23}&\frac{4 n^2\left(c^2-2 n^2\right)}{c^3 R^2 \omega_n^2}&\frac{1}{J}\left( 8 n^2 \pi ^2 \lambda '-96 n^4 \pi ^4 \lambda'^2 +768 n^6 \pi^6 \lambda'^3
-6144 n^8 \pi ^8 \lambda'^4+\dots\right)\\ \hline
u_{24}&\frac{4 n^2\left(c^2-2 n^2\right)}{c^3 R^2 \omega_n^2}&\frac{1}{J}\left( 8 n^2 \pi ^2 \lambda '-96 n^4 \pi ^4 \lambda'^2 +768 n^6 \pi^6 \lambda'^3
-6144 n^8 \pi ^8 \lambda'^4+\dots\right)\\ \hline
\end{array}
\end{equation*}\end{table}
The Tables above show that each energy has an even, at least
double, multiplicity, as one expects from the symmetry of the
Bethe framework configurations, as we shall discuss in the Section
below. Thus the above string spectrum, one of the main results of
this Paper, can be consistently compared with the solutions of the
Bethe equations, providing a significant test of them and of the
integrability of the string sigma model in the near-BMN limit.
Notice also that the results above, together with the eight
bosonic states addressed in~\cite{Astolfi:2011ju}, sheds a
complete light on the spectrum of all the 32 bosonic two-impurity
light states of the theory. Finally, for comparison with the Bethe
Ansatz framework, the string spectrum above, exact in $\lambda'$,
can be expanded in power series, as shown in the Table \ref{cts}
below. The $c_i$ are the coefficients of the $\lambda'$ expansion
of the spectrum defined as
\beq\label{expn} \epsilon=\epsilon_0 + \frac{1}{J}\sum_{i=1}
c_i\lambda'^i \left(8 \pi^2 n^2\right)^i
+\mathcal{O}\left(\frac{1}{J^2}\right),\eeq
where $\epsilon_0$ is the term of order $1/J^0$, $n$ is the mode
number of the two-oscillator state.
\begin{table}[h!]\caption{\label{cts} Coefficients
 $c_i$ and state multiplicities.} \beq
\begin{array}
{|c|c|c|c|c|c|c|}\hline\hline \mbox{State type (BB,FF) and
Nr.$i$ } &c_1&c_2&c_3&c_4&c_5&\mbox{Multiplicity}\\
\hline\hline FF: 23,24 &1 &
-\frac{3}{2} & \frac{3}{2} & -\frac{3}{2} & \frac{3}{2} &2_{FF}\\
\hline BB: 5,7; FF: 9,10,17,18,19,20,21,22  &0 & -\frac{1}{2} &
\frac{1}{2} & -\frac{1}{2} & \frac{1}{2} & 10=8_{FF}+2_{BB} \\
\hline BB: 2,3,6,8; FF: 11,12,13,14,15,16 &-1 & \frac{1}{2} &
-\frac{1}{2} & \frac{1}{2} & -\frac{1}{2} &10=6_{FF}+4_{BB} \\
\hline BB: 1,4 &-2 &
\frac{3}{2} & -\frac{3}{2} & \frac{3}{2} & -\frac{3}{2}& 2_{BB}\\
\hline\hline
\end{array}
\eeq\end{table}

\section{Energies of Bethe states}
\label{sec:Bethe}
The Bethe roots are quantized through the algebraic equations~\cite{Gromov:2008qe}:
\begin{eqnarray} \label{bethe}
1&=&
\prod_{j=1}^{K_2}
\frac{u_{1,k}-u_{2,j}+\frac{i}{2} }{u_{1,k}-u_{2,j}-\frac{i}{2} }
\prod_{j=1}^{K_{4}}
\frac{1-1/x_{1,k} x^+_{4,j}}{1-1/x_{1,k}x_{4,j}^-}
\prod_{j=1}^{K_{\bar 4}}
\frac{1-1/x_{1,k} x^+_{\bar 4,j}}{1-1/x_{1,k}x_{\bar 4,j}^-} \,,
 \nn \\
\nn 1&=& \prod_{j\neq k}^{K_2} \frac{u_{2,k}-u_{2,j}-i
}{u_{2,k}-u_{2,j}+i } \prod_{j=1}^{K_1}
\frac{u_{2,k}-u_{1,j}+\frac{i}{2} }{u_{2,k}-u_{1,j}-\frac{i}{2} }
\prod_{j=1}^{K_3} \frac{u_{2,k}-u_{3,j}+\frac{i}{2}
}{u_{2,k}-u_{3,j}-\frac{i}{2} }\,,
\\
\nn 1&=&
\prod_{j=1}^{K_2}
\frac{u_{3,k}-u_{2,j}+\frac{i}{2} }{u_{3,k}-u_{2,j}-\frac{i}{2} }
\prod_{j=1}^{K_4}
\frac{x_{3,k} -x^+_{4,j}}{x_{3,k} -x^-_{4,j}}
\prod_{j=1}^{K_{\bar 4}}
\frac{x_{3,k} -x^+_{\bar 4,j}}{x_{3,k} -x^-_{\bar 4,j}} \,
\\
\left(\frac{x^+_{4,k}}{x^-_{4,k}}\right)^{
L} &=&
\prod_{j\neq k}^{K_4}
\frac{u_{4,k}-u_{4,j}+i}{u_{4,k}-u_{4,j}-i}  \,
\prod_{j=1}^{K_1}
\frac{1-1/x^-_{4,k} x_{1,j}}{1-1/x^+_{4,k} x_{1,j}}
\prod_{j=1}^{K_3}
\frac{x^-_{4,k}-x_{3,j} }{x^+_{4,k}-x_{3,j}}   \times
 \\
\nn &\times &\prod_{j=1}^{K_4}
\sigma_{\rm BES}(u_{ 4,k},u_{ 4,j})   \prod_{j=1}^{K_{\bar 4}}  \sigma_{\rm BES}(u_{ 4,k},u_{ \bar 4,j})  \,, \\
\left(\frac{x^+_{\bar 4,k}}{x^-_{\bar  4,k}}\right)^{
L} &=&
\prod_{j=1}^{K_{\bar 4}}
\frac{u_{\bar  4,k}-u_{\bar 4,j}+i}{u_{\bar 4,k}-u_{\bar 4,j}-i}  \,
\prod_{j=1}^{K_1}
\frac{1-1/x^-_{\bar 4,k} x_{1,j}}{1-1/x^+_{\bar 4,k} x_{1,j}}
\prod_{j=1}^{K_3}
\frac{x^-_{\bar 4,k}-x_{3,j} }{x^+_{\bar 4,k}-x_{3,j}}   \times
\nn \\
&\times &\prod_{j\neq k}^{K_{\bar 4}}
\sigma_{\rm BES}(u_{ \bar 4,k},u_{ \bar 4,j}) \prod_{j=1}^{K_{  4}} \sigma_{\rm BES}(u_{ \bar 4,k},u_{ 4,j}) \nn \,,
\end{eqnarray}
where the spectrum of string energies is expressed in terms of the
roots $u_4$ and $u_{\bar 4}$, which carry momentum, as follows:
\begin{equation}
E =\ff{\cal Q}_2 \,,\la{eq:E}
\end{equation}
the conserved charges being expressed in terms of the roots as
\begin{equation}
{\cal Q}_n= \sum_{j=1}^{K_4} \textbf{q}_n(u_{4,j})+
\sum_{j=1}^{K_4} \textbf{q}_n(u_{\bar 4,j}) \,\,\, , \,\,\,
\textbf{q}_n=\frac{i}{n-1}
\left(\frac{1}{(x^+)^{n-1}}-\frac{1}{(x^-)^{n-1}}\right) \,.
\la{charges}
\end{equation}
The Zhukovsky variables are  defined in terms of the roots as
\begin{equation}
x+\frac{1}{x}= \frac{u}{\ff}\;\;,\;\;x^\pm+
\frac{1}{x^\pm}=\frac{1}{\ff}\left(u\pm\frac i2\right) \,.
\end{equation}
Recalling that $p_j=\frac{1}{i}\log\frac{x^+_{4,j}}{x^-_{4,j}}$
and $\bar p_j=\frac{1}{i}\log\frac{x^+_{\bar 4,j}}{x^-_{\bar
4,j}}$, we have
\begin{equation} \label{disprelbethe}
E =\sum_{j=1}^{K_4} \frac{1}{2}\left(\sqrt{1+16 \ff^2 \sin^2
\frac{p_j}{2}} -1\right)+\sum_{j=1}^{K_{\bar 4}}
\frac{1}{2}\left(\sqrt{1+16 \ff^2 \sin^2 \frac{\bar p_j}{2}}
-1\right) \,.
\end{equation}
At large 't Hooft coupling we have
\begin{equation}
\ff\simeq \sqrt{\lambda/2} \,.
\end{equation}
The rapidity variable expressed  in terms of the momentum of the
roots is given by
\begin{equation}
u_{4,j}=\frac{1}{2}\cot{\left(\frac{p_j}{2}\right)}\sqrt{1
+16\ff^2 \sin{\left(\frac{p_j}{2}\right)}^2}\,.
\end{equation}
In the near plane wave limit, the BES
kernel~\cite{Beisert:2006ez} reduces to the AFS phase
factor~\cite{Arutyunov:2004vx}:
\begin{equation}
\sigma_{\rm AFS}(u_j,u_k)=e^{i\theta_{jk}}\,,
\end{equation}
where
\begin{equation}
\theta_{jk}=\sum_{r=2}^\infty\ff
\left[\textbf{q}_r(x_j)\textbf{q}_{r+1}(x_k)-
\textbf{q}_r(x_k)\textbf{q}_{r+1}(x_j)\right]\,.
\end{equation}
The Bethe equations can be solved for the momenta  $p_j$ in the
near plane wave limit only with a judicious perturbative Ansatz as
\begin{equation} \label{exp}
p_j= \frac{2 \pi n_j}{J}  +\frac{A}{J^2}+\frac{B
\lambda'}{J^2}+\frac{C \lambda'^2}{J^2}+\frac{D
\lambda'^3}{J^2}\dots\,,
\end{equation}
where we solve order by order determining the expansion coefficients. Eventually we plug the solution for the momenta in
the dispersion relation \eqref{disprelbethe} to get the spectrum.

\subsection{Warm up: recap  of the $SU(2)\times SU(2)$ subsector}

Consider the cases $(K_{u_4}, K_{u_{\bar 4}}, K_{u_1}, K_{u_2},
K_{u_3})=(2,0,0,0,0)$ and $(K_{u_4}, K_{u_{\bar 4}}, K_{u_1},
K_{u_2}, K_{u_3})=(0,2,0,0,0)$, which are clearly identical . Due
to the level matching  condition, we have only one independent
momentum, $p$. Plugging the expansion \eqref{exp} in the Bethe
equations, one gets, up to order $\lambda'^2$ and $\frac{1}{J}$:
\begin{equation}
\frac{1}{J}\left[A- 2 \pi n+\lambda' \left(B+8 n^3 \pi^3\right)+
\lambda'^2\left(C-32 n^5 \pi^5\right)+\lambda'^3\left(D+192 n^7
\pi^7\right)\right]=0\,,
\end{equation}
which completely determines the momentum up to the desired perturbative order.
We have
\begin{equation}
A= 2 n \pi ,\;\;\; B =-8 n^3 \pi^3,\;\;\;C=32 n^5 \pi^5\,,D=-192
n^7\pi^7,
\end{equation}
which plugged in the dispersion  relation \eqref{disprelbethe} gives the
spectrum:
\begin{equation}\label{e20000}
E_{20000}=4 n^2 \pi^2 \lambda'-8 n^4 \pi^4 \lambda'^2+32 n^6 \pi^6
\lambda'^3+\frac{1}{J}\left(8 n^2  \pi^2 \lambda'-64 n^4 \pi^4
\lambda'^2+448 n^6 \pi^6 \lambda'^3 - 3328 n^8 \pi^8
\lambda'^4\right)\dots
\end{equation}
which is the spectrum of the string states $| s_{1,2} \rangle =
\left(a^{1,2}_n\right)^\dagger\left(a^{1,2}_{-n}\right)^\dagger  |
0 \rangle$, addressed in~\cite{Astolfi:2008ji}.

Consider now the case $(K_{u_4}, K_{u_{\bar 4}}, K_{u_1}, K_{u_2},
K_{u_3})=(1,1,0,0,0)$. Similarly, due to the level matching condition, there
is only one independent momentum, $p$. Yet we can build two
different configurations, which shall be degenerate: the $u_4$
root carrying momentum $p$ and the $u_{\bar 4}$ carrying $-p$, or
viceversa. The perturbative expansion of the Bethe equations
reads:
\begin{equation}
\frac{1}{J}\left[A+\lambda'  B+\lambda'^2\left(C+16 n^5
\pi^5\right)+\lambda'^4\left(D+128 n^7 \pi^7\right) \right]=0\,,
\end{equation}
which gives
\begin{equation}
A= 0 ,\;\;\; B =0,\;\;\;C=-16 n^5 \pi^5\,, D=-128n^7\pi^7,
\end{equation}
and therefore
\begin{equation}\label{e11000}
E_{11000}=4 n^2 \pi^2 \lambda'-8 n^4 \pi^4 \lambda'^2+32 n^6 \pi^6
\lambda'^3-\frac{1}{J}\left(64 n^6 \pi^6 \lambda'^3-768 n^8\pi^8
\lambda'^4\right)+\dots\,,
\end{equation}
which is the spectrum of the  string states $| t_{1,2} \rangle =
\left(a^{1,2}_n\right)^\dagger\left(a^{2,1}_{-n}\right)^\dagger |
0 \rangle$, addressed in~\cite{Astolfi:2008ji}.

Actually the matching, which can be perturbatively checked at arbitrary high orders in $\lambda'$, between these solutions to the Bethe equations and the near plane wave spectrum of string states in the $SU(2)\times SU(2)$ subsector, discussed in~\cite{Astolfi:2008ji}, has provided one of the earliest tests of the all loop asymptotic Bethe Ansatz proposed in~\cite{Gromov:2008qe}.

\subsection{How to deal with auxiliary roots}

The one-magnon bosonic states  correspond to the configurations:
$(K_{u_4}, K_{u_{\bar 4}}, K_{u_1}, K_{u_2},
K_{u_3})=(1,0,0,0,0)$, $(0,1,0,0,0)$, $(1,0,1,1,1)$ and
$(1,0,1,1,1)$. The one-magnon fermionic configurations are instead
$(K_{u_4}, K_{u_{\bar 4}}, K_{u_1}, K_{u_2},
K_{u_3})=(1,0,1,0,0)$, $(0,1,1,0,0)$, $(1,0,1,1,0)$,
$(0,1,1,1,0)$. Out of these, 32 two-magnon states may be formed.
We are interested in those having degenerate energies in the
plane-wave limit, since they correspond to the string
configurations we have studied in Section~\ref{sec:stringcomp}. In
the boson-boson sector these are $(K_{u_4}, K_{u_{\bar 4}},
K_{u_1}, K_{u_2}, K_{u_3})=(1,1,1,1,1)$, $(2,0,1,1,1)$ and
$(0,2,1,1,1)$; in the fermion-sector these are $(K_{u_4},
K_{u_{\bar 4}}, K_{u_1}, K_{u_2}, K_{u_3})=(1,1,2,0,0)$,
$(1,1,2,1,0)$, $(1,1,2,2,0)$, $(2,0,2,0,0)$, $(2,0,2,1,0)$,
$(2,0,2,2,0)$, $(0,2,2,0,0)$, $(0,2,2,1,0)$ and $(0,2,2,2,0)$.

Taking into account the exact $\mathbb{Z}_2$
degeneracy due to $p\to -p$ symmetry and the double occurrence of
$(..210)$ and $(..111)$ states due to branching of auxiliary roots
we obtain 24 states having plane-wave degenerate spectrum, which
exactly corresponds to the degenerate string two oscillator
spectrum.

Below we therefore solve Bethe equations for these states
and find their spectrum. We work at large $\lambda$, in the first
order in $\frac{1}{J}$, and up to the $4$th order in
$\lambda'=\frac{\lambda}{J^2}$. The procedure is a perturbative expansion in $\frac{1}{J}$ and $\lambda'$, parallel to the warm up exercise of the $SU(2)\times SU(2)$ recalled above.

The order in $\lambda'$ seems to be improvable {\it ad infinitum};
we chose the fourth order due to two considerations. First, it is
order $\lambda'^3$ where discrepancy between the spectrum of gauge
invariant operators and near plane-wave string energies has been
first found for the $\rm{AdS}_5/\rm{CFT}_4$ correspondence, and
cured with the introduction of the AFS phase
factor~\cite{Arutyunov:2004vx}  interpolating between weak and
strong coupling. Thus, this feature having been substantially
inherited in the $\rm{AdS}_4/\rm{CFT}_3$ correspondence, as one
can read from the Bethe equations~\eqref{bethe}, agreement at
$\lambda'^4$ is such a nontrivial statement that be considered an
all-order result. Second, larger values of the order become
problematic (but not impossible) on {\it Mathematica}.

To regularize Bethe equations for those solutions carrying
auxiliary roots $u_i=0,u_i=\infty$ perturbatively, one should add
twist parameters $\epsilon_{1,2,3}$, as suggested
in\mycite{Volin:2010xz}, in the following way:

\begin{eqnarray} \label{bethe1}
e^{i\epsilon_1}&=& \prod_{j=1}^{K_2}
\frac{u_{1,k}-u_{2,j}+\frac{i}{2} }{u_{1,k}-u_{2,j}-\frac{i}{2} }
\prod_{j=1}^{K_{4}} \frac{1-1/x_{1,k}
x^+_{4,j}}{1-1/x_{1,k}x_{4,j}^-} \prod_{j=1}^{K_{\bar 4}}
\frac{1-1/x_{1,k} x^+_{\bar 4,j}}{1-1/x_{1,k}x_{\bar 4,j}^-} \,,
 \nn \\
\nn e^{i\epsilon_2}&=& \prod_{j\neq k}^{K_2}
\frac{u_{2,k}-u_{2,j}-i }{u_{2,k}-u_{2,j}+i } \prod_{j=1}^{K_1}
\frac{u_{2,k}-u_{1,j}+\frac{i}{2} }{u_{2,k}-u_{1,j}-\frac{i}{2} }
\prod_{j=1}^{K_3} \frac{u_{2,k}-u_{3,j}+\frac{i}{2}
}{u_{2,k}-u_{3,j}-\frac{i}{2} }\,,
\\
\nn e^{i\epsilon_3}&=& \prod_{j=1}^{K_2}
\frac{u_{3,k}-u_{2,j}+\frac{i}{2} }{u_{3,k}-u_{2,j}-\frac{i}{2} }
\prod_{j=1}^{K_4} \frac{x_{3,k} -x^+_{4,j}}{x_{3,k} -x^-_{4,j}}
\prod_{j=1}^{K_{\bar 4}} \frac{x_{3,k} -x^+_{\bar 4,j}}{x_{3,k}
-x^-_{\bar 4,j}} \, ,
\end{eqnarray}

The Bethe equations must be solved perturbatively for the momenta
of the physical roots, in terms of the solution for the auxiliary
roots expressed through the parameters  $\epsilon_{1,2,3}$. At the
end of the procedure, one takes the limit $\epsilon_{1,2,3}\to 0$
and plugs the solution for the momenta in the dispersion
relation~\eqref{disprelbethe}. The spectrum of the boson-boson
configurations, up to order $\lambda'^4$, is given in the
following Table\myref{bbba}. In the table we show the quantity $
\varepsilon$, defined as
\beq
E=\varepsilon_0+\frac{\varepsilon}{J}+O\left(\frac{1}{J^2}\right).\eeq
In the ``Note'' column we show the final form of the lowest Bethe
equation (the one for $u_4$) that is being actually solved. The
phase $\sigma_{AFS}$ and spectral variable $u(p)$ is meant as
function of $p$, the latter given by\myref{exp}. It is important
to realize that these energy corrections are quite different from
those for $20000$, $11000$ states\myref{e11000},\myref{e20000},
despite the singularity of the roots. Auxiliary roots going to
infinity (in the $u$ plane) do not result in full cancellation of
their respective contributions in the equations for $u_4$ and
$u_{\bar{4}}$, since the $x(u)$ are different for these solutions.
A solution of $20111$ type with $x_{1,2,3}=\infty$ would have been
equivalent to $20000$; however, in our case
$x_1=\infty,x_2=\infty,x_3=0$ which yields a solution of a totally
different type due to extra $x^+/x^-$ factor coming from the
right-hand side of Bethe equation.
\begin{table}[h!]
\caption{\label{bbba}Boson-boson spectrum from Bethe Ansatz}
\hspace{-1cm}\begin{equation*}
\begin{array}{|ccccc|rl|l|l|}\hline
\multicolumn{5}{|c|}{\mbox{state}}
&\multicolumn{2}{|c|}{\mbox{Energy coefficient $\varepsilon$}}
&\multicolumn{1}{|c|}{\mbox{Auxiliary roots}}& \mbox{Note}\\
\hline
K_4&K_{\bar{4}}&K_3&K_2&K_1&&&&\\
\hline\hline 1&1&1&1&1 & -8 n^2 \pi ^2 \lambda '&+32 n^4 \pi ^4
\lambda'^2-&u_1= \frac{1}{\epsilon _1},x_1\to \infty& e^{i p(J+1)}
=\sigma_{AFS}\\
&&&&&-256 n^6 \pi ^6 \lambda'^3& +2048 n^8 \pi ^8 \lambda'^4&
u_2=\frac{1}{\epsilon_2},x_2\to \infty &\mbox{$J+1$ due to extra
$x^-/x^+$ }
\\&&&&&&&  u_3=\frac{1}{\epsilon_3},x_3\to 0&\mbox{from
$\prod^{K_3}\left(\cdots\right )$} \\
\hline 
2&0&1&1&1& &-32 n^4 \pi ^4 \lambda'^2-&
\multicolumn{1}{|c|}{\mbox{\rule{1cm}{1pt}
\slash\slash\rule{1cm}{1pt}}}& e^{i p(J+1)}=\frac{2u+i}{2u-i}\sigma_{AFS}\\

&&&&&+256 n^6 \pi ^6 \lambda'^3&-2048 n^8 \pi ^8 \lambda'^4 &&\\

\hline\hline
1&1&1&1&1 & -16 n^2 \pi ^2 \lambda '&+ 96 n^4 \pi ^4  \lambda'^2-&
u_1= \frac{1}{\epsilon _1}&
e^{ipJ}=\frac{J+8 i \pi ^3 \lambda'  n^3-4 i \pi  n}{J}\sigma_{AFS}\\
&&&&&&& u_2= \frac{1}{\epsilon_2}&
\\&&&&&
-768 n^6  \pi^6 \lambda'^3&
+6144 n^8 \pi ^8 \lambda'^4&
 x_3=\frac{1}{2 \sqrt{2} n \pi  \sqrt{\lambda '}}&\\
\hline 
2&0&1&1&1&  -8 n^2 \pi ^2 \lambda '&+32 n^4 \pi ^4
\lambda'^2-&\multicolumn{1}{|c|}{\mbox{\rule{1cm}{1pt}
\slash\slash\rule{1cm}{1pt}}}& e^{ipJ}=\frac{2u+i}{2u-i}\frac{J+8
i \pi ^3 \lambda'  n^3-4 i
\pi  n}{J}\sigma_{AFS}\\
&&&&&-256 n^6 \pi ^6 \lambda'^3& +2048 n^8 \pi ^8 \lambda'^4 &&\\
\hline\hline
\end{array}
\end{equation*}
\end{table}

Similarly in the next Table\myref{ffba} below we show the
bifermionic part of the two-magnon sector of Bethe Ansatz,
$\varepsilon$ defined as above.
\begin{table}[h!]\caption{\label{ffba}
Fermion-fermion spectrum from Bethe Ansatz}
\begin{equation*}
\begin{array}{|ccccc|rl|l|}\hline
\multicolumn{5}{|c|}{\mbox{state}}
&\multicolumn{2}{|c|}{\mbox{Energy coefficient $\varepsilon$}}
&\multicolumn{1}{|c|}{\mbox{Auxiliary roots}}\\
K_4&K_{\bar{4}}&K_3&K_2&K_1&&&\\
\hline\hline 1&1&2&2&0 & &-32 n^4 \pi ^4 \lambda'^2+&u_{21}=
\frac{1}{2} \left(u_{31}+u_{32} -\right.\\ &&&&&+256 n^6 \pi ^6
\lambda'^3&-2048 n^8 \pi ^8 \lambda'^4 & \left.-i \sqrt{u_{31}^2-2
u_{32} u_{31} +u_{32}^2+1}\right)\\ &&&&&&& u_{22}=\frac{1}{2}
\left(u_{31}+u_{32}+\right.\\
&&&&&&& \left.+i \sqrt{u_{31}^2-2 u_{32} u_{31}
+u_{32}^2+1}\right)\\&&&&&&& x_{31}= -2 i-\frac{i}{2 n^2 \pi ^2
\lambda '}+2 i n^2 \pi ^2 \lambda '\\&&&&&&& x_{32}=\,\, 2
i+\frac{i}{2 n^2 \pi ^2 \lambda '}-2
i n^2 \pi ^2 \lambda ' \\
\hline
%
2&0&2&2&0& 8 n^2 \pi ^2 \lambda '&-96 n^4 \pi ^4 \lambda'^2+
&\multicolumn{1}{|c|}{\mbox{\rule{1cm}{1pt}\slash\slash\rule{1cm}{1pt}}}\\
&&&&&+768 n^6 \pi ^6 \lambda'^3&-6144 n^8 \pi ^8 \lambda'^4&\\
\hline\hline
%
1&1&2&1&0 &  -8 n^2 \pi ^2 \lambda '&+32 n^4 \pi ^4 \lambda'^2-&
u_2=\frac{1}{2} \left(u_{31}+u_{32}\right)\\&&&&&-256 n^6 \pi ^6
\lambda'^3&+2048 n^8 \pi ^8 \lambda'^4& x_{31},x_{32}\,\,
\mbox{solutions of}\\&&&&&&&
\,\,\frac{(x_{3}-x(u+i/2))(x_{3}-x(-u+i/2))}{(x_{3}-x(u-i/2))
(x_{3}-x(-u-i/2))}=e^{i\epsilon_3}
\\
\hline 
2&0&2&1&0& &-32 n^4 \pi ^4 \lambda'^2+&
\multicolumn{1}{|c|}{\mbox{\rule{1cm}{1pt}\slash\slash\rule{1cm}{1pt}}}\\
&&&&&+256 n^6 \pi ^6 \lambda'^3&-2048 n^8 \pi ^8 \lambda'^4 &\\
\hline\hline
%
1&1&2&1&0 &  -8 n^2 \pi ^2 \lambda '&+32 n^4 \pi ^4 \lambda'^2+&
\, u_2= \frac{1}{2} \left(u_{31}+u_{32}+\frac{4}{\epsilon
_2}\right)\\&&&&&-256 n^6 \pi ^6 \lambda'^3&+2048 n^8 \pi ^8
\lambda'^4& x_{31}= \frac{\left(1+4 u(p)^2\right) \epsilon _3}{4
\sqrt{2} J \sqrt{\lambda '}},\\  &&&&&&& u(p)=\frac{1}{2} \cot
\left(\frac{p}{2}\right) \sqrt{1+2\lambda
\sin^2\frac{p}{2}}\\&&&&&&& x_{32}= -\frac{\left(1+4
u(p)^2\right)  \left(-2i+\epsilon _3\right)}{4 \sqrt{2} J \sqrt{\lambda '}}\\
\hline 
2&0&2&1&0& &-32 n^4 \pi ^4 \lambda'^2+&
\multicolumn{1}{|c|}{\mbox{\rule{1cm}{1pt}\slash\slash\rule{1cm}{1pt}}}\\
&&&&&+256 n^6 \pi ^6 \lambda'^3&-2048 n^8 \pi ^8 \lambda'^4 &\\
\hline\hline
%
1&1&2&0&0 &  -8 n^2 \pi ^2 \lambda '&+32 n^4 \pi ^4 \lambda'^2-&
x_{31}=\frac{J \epsilon_3 }{4 \sqrt{2} n^2 \pi ^2 \sqrt{\lambda
'}}\\&&&&&-256 n^6 \pi ^6 \lambda'^3&+2048 n^8 \pi ^8 \lambda'^4&
x_{32}=\frac{J
\epsilon_3 }{4 \sqrt{2} n^2 \pi ^2 \sqrt{\lambda '}}\\
\hline 
2&0&2&0&0& &-32 n^4 \pi ^4 \lambda'^2+
&\multicolumn{1}{|c|}{\mbox{\rule{1cm}{1pt}
\slash\slash\rule{1cm}{1pt}}}\\
&&&&&+256 n^6 \pi ^6 \lambda'^3&-2048 n^8 \pi ^8 \lambda'^4&\\
\hline\hline
\end{array}
\end{equation*}
\end{table}

This table is quite remarkable, since all states presented here
are also found on the string side, and the energies coincide up to
the highest order done on the Bethe Ansatz side. Given this
coincidence, as discussed above, this clearly points to an
all-order equivalence for the finite-size corrections calculated
from the  Bethe Ansatz and from the string theory, in the limit
$\lambda'\to 0$, $J\to\infty$, for all the two impurity light
bosonic states. This is a remarkable further test of the Bethe
Ansatz framework and a significant effort towards  quantum
integrability of strings in $AdS_4\times \mathbb{CP}_3$.

To facilitate this comparison and summarize, let us represent the
spectrum in a more concise form, showing the expansion
coefficients in powers of $\lp$ and the multiplicities of the
states. For conciseness we do not write out the states $02...$,
since they are fully equivalent to the corresponding states
$20...$. All states $11...$ are twice degenerate to all orders due
to $n\to -n$ symmetry. Thus each of the states on the right hand
side must be duplicated, which yields correct matching of the
number of the degrees of freedom. Boson-boson sector is compared
in the Table\myref{bbstba}.
\begin{table}[h!]\caption{\label{bbstba}Boson-boson spectrum comparison}
\beq
\begin{array}{|lllll|c|ccccl|l|}\hline
\multicolumn{5}{|c|}{\mbox{Expansion
coefficient}}&\mbox{Multiplicity}&\multicolumn{5}{|c|}{\mbox{Corresponding
BA states}}&\mbox{Corresponding ST states}\\ \hline
\,\,\,c_1&\,\,\,c_2&\,\,\,\,c_3&\,\,\,\,c_4&\,\,\,\,c_5&
&K_4&K_{\bar{4}}&K_3&K_2&K_1&\mbox{State nr.}\\
\hline\hline \hphantom{-} 0 & -\frac{1}{2} &
\hphantom{-}\frac{1}{2} &
-\frac{1}{2} & \hphantom{-}\frac{1}{2}&2&
2&0&1&1&1_{\mbox{branch 1}}& 5,7\\
\hline\hline
-1 & \hphantom{-}\frac{1}{2} & -\frac{1}{2} &
\hphantom{-}\frac{1}{2} & -\frac{1}{2}
&4& 2&0&1&1&1_{\mbox{branch 2}} &2,3,6,8\\
&&&&&&1&1&1&1&1_{\mbox{branch 1}}& \\
\hline\hline -2 & \hphantom{-}\frac{3}{2} & -\frac{3}{2} &
\hphantom{-}\frac{3}{2} & -\frac{3}{2}
&2& 1&1&1&1&1_{\mbox{branch 2}}& 1,4\\
\hline\hline
\end{array}
\eeq
\end{table}
Analogously, fermion-fermion spectrum comparison is done in the
Table\myref{ffstba}.
\begin{table}[h!]\caption{\label{ffstba}
Fermion-fermion spectrum comparison}
\beq
\begin{array}{|lllll|c|ccccl|l|}\hline
\multicolumn{5}{|c|}{\mbox{Expansion
coefficient}}&\mbox{Multiplicity}&
\multicolumn{5}{|c|}{\mbox{Corresponding BA
states}}&\mbox{Corresponding ST states}\\ \hline
\,\,\,c_1&\,\,\,c_2&\,\,\,\,c_3&\,\,\,\,c_4&\,\,\,\,c_5&
&K_4&K_{\bar{4}}&K_3&K_2&K_1&\mbox{State nr.}\\
\hline\hline \hphantom{-}1 & -\frac{3}{2} &
\hphantom{-}\frac{3}{2} & -\frac{3}{2} & \hphantom{-}\frac{3}{2}&
 2& 2&0&2&2&0&23,24\\ \hline\hline
\hphantom{-}0 & -\frac{1}{2} &  \hphantom{-}\frac{1}{2} &
-\frac{1}{2} & \hphantom{-}\frac{1}{2} &
8& 1&1&2&2&0& 9, 10, 17, 18, 19, 20, 21, 22\\
&&&&&& 2&0&2&1&0_{\mbox{branch 1}}&\\
&&&&&& 2&0&2&1&0_{\mbox{branch 2}}&\\
&&&&&& 2&0&2&0&0&\\ \hline\hline -1 & \hphantom{-}\frac{1}{2} &
-\frac{1}{2} & \hphantom{-}\frac{1}{2} & -\frac{1}{2}&
6&1&1&2&1&0_{\mbox{branch 1}}& 11, 12, 13, 14, 15, 16\\
&&&&&& 1&1&2&1&0_{\mbox{branch 2}}&\\
&&&&&& 1&1&2&0&0&\\
\hline\hline
\end{array}
\eeq
\end{table}

\subsection{Claim to exactness}
The spectacular coincidence of the $\lambda'$ expansions for Bethe
energies with the string energies supposes that it might  be
exact. This exactness can actually be seen directly in some of the
cases. In the previous subsection the procedure to solve Bethe
equations was to start with ``highest'' auxiliary nodes $1,2,3$,
then descend to the physical magnons $4,\bar{4}$. Here we act
reversely: start with the physical node, the momentum of which is
known exactly in $\lambda'$  form the exact string spectrum

\beq \epsilon=\frac{4 \pi ^2 n^2 \lambda ' \left(A-\frac{8 \pi ^2
(A+1) n^2 \lambda '}{8\pi ^2 n^2 \lambda '+1}\right)}{J},\eeq
where $A=2,0,-2,-4$ for the four admissible energy values of our
spectrum. We can thus use the highest auxiliary node equation as a
test. We have seen for several states of Bethe Ansatz (e.g. the
$(1,1,1,1,1),\, (2,0,1,1,1)$ states) that the first equation is
non-trivially satisfied in a regular manner, that is, by a
systematic improvement of the expansion one can satisfy the Bethe
equation up to all orders.

\section{Conclusion}
\label{sec:conclusions}
\subsection{Summary of the near BMN calculations}
The main results of this Paper can be summarized as follows:
\begin{itemize}
\item Our calculations provide a highly non-trivial test for the
validity of the string Hamiltonian for three and four-particle
interaction vertices in a near Penrose limit computed
in~\cite{Astolfi:2009qh}.

\item The one-loop correction to the single-magnon dispersion
relation, as expected, is the same for bosonic and fermionic
excitation, it is finite and exponentially small in $J$ for $J$ large.  The
regularization prescription implied by the cubic Hamiltonian and
by consequent unitarity arguments, gives a vanishing one loop
correction to the strong-weak coupling interpolating function
$h(\lambda)$, $a_1=0$.

\item In the two-particle sector the finite-size corrections (the
$1/J$-corrections) to magnon interaction energies on the string
side and on the Bethe-Ansatz side are exactly the same up to
the fourth order in $\lp\equiv \frac{\lambda}{J^2}$.
\end{itemize}

The second result in the one-particle sector relies on the
argument about the unitarity-preserving property of the
regularization. This argument, in view of its very general nature,
should be applicable to all Bethe Ansatz states, yet it remains an
open problem e.g. how exactly it would work for e.g. the GKP
case~\cite{Gubser:1998bc}, or spinning string states. However we
may so far claim that a possible reason of the problems arising
with the equal-frequency regularization (linear divergencies;
disagreement between strings and the algebraic curve) is a
possible unitarity violation by regularization. Thus the problems
arising with it may be, somewhat loosely, called ``unitarity
anomaly''.

The significance of the third result is to establish another
instance of the ``mutual understanding'' between the conjectured BA at
all couplings with strings on $AdS_4\times \C P^3$ in Penrose
limit. The BA is asymptotic and thus is not {\it a priori}
expected to work at strong coupling and arbitrary length.  Yet it
works, as established by our third result, not just
asymptotically at $J\to\infty$ but also at least at the order
$1/J$. And at that order it is completely non-trivial that
equivalence between the spectrum states holds at the fourth order
in $\lambda'$.

We conjecture that the equivalence is actually an exact one, and
extends towards higher orders in $\frac{1}{J}$.

\subsection{Further questions}

The full Bethe Ansatz, with {\it all} finite size and loop
corrections, at both weak and strong coupling is encoded in the
Y-system\mycite{Gromov:2009at,Bombardelli:2009xz}. For
$AdS_5/CFT_4$ this infinite system of functional equations has
recently been shown to be equivalent to the T-system in
\mycite{Gromov:2011cx}, which is reducible to a finite number of
integral equations. Thus of importance would be to test the
results for finite-size corrections in strings against the
T-system.  The L\"uscher terms are absent in the asymptotic Bethe
Ansatz; string calculations would normally see them directly; in
our case the exponentially-suppressed finite-size corrections look
precisely like the typical L\"uscher corrections do. Of extreme
interest would be to compare in the one-loop sector the T-system
with the direct L\"uscher calculations (for a review
see\mycite{Janik:2010kd} for example), and with our string
calculation of the finite size correction to the dispersion
relation.

It is the strong coupling limit where  the Y-system calculation
should be easier to perform, since for strong coupling the
functional/integral equations become algebraic and a full analytic
solution becomes possible. Yet to our knowledge these solutions
have so far been applied to  GKP~\cite{Gubser:1998bc} states
mostly, and not to BMN. Therefore this should be one of the major
lines of further research - to obtain the self-energy finite-size
corrections for the near-BMN spectrum directly from Y or T system,
comparing them with the string calculation.

Probably the most urgent and straightforward  further direction of
the present work is, on the grounds of the same techniques, its
extension to the computation of the finite size corrections of
string states involving at least one heavy mode. Actually the
Penrose limit of the geometry decouples light and heavy modes,
having different dispersion relation. Yet they look equally
fundamental, both being described by a Fourier series of
harmonic-oscillator like modes, such that one might think that the
fundamental degrees of freedom of the theory are 8B + 8F. When
dealing with finite size corrections, we have extensively
discussed how infinite sums appear in the computation of the
spectrum, which need to be regularized. The momentum conservation
at the cubic light-light-heavy vertex forces the cutoff on the
mode numbers of a heavy mode being twice as that of a light one.
This is the first glimpse of the interpretation of a heavy mode
being, rather than fundamental, a bound state of two light modes,
such that indeed the basic degrees of freedom are 4B+4F, as in the
Bethe Ansatz framework. Actually the Bethe program gives a recipe
for building a single light heavy oscillator state, being a
composite of the fundamental roots. If this picture is correct,
the finite size spectrum of light-heavy or heavy-heavy two
oscillator states must match the corresponding solutions of the
Bethe equations. If this occurs, the puzzle about the
interpretation of the heavy mode would be definitely solved.

The test of the $AdS_5$ results at strong coupling has taken place
at two loops for GKP strings, and at 1 loop for the near-BMN
limit, yielding perfect agreement to the Y-system. Therefore the
second-loop corrections to the self-energies and the first-loop
corrections to scattering amplitudes could be interesting to
calculate. Our Hamiltonian  approach here would be extremely hard
to implement, so we guess that perhaps the continuous Lagrangian
field-theoretical approach~\cite{Abbott:2011xp} might be used
provided it is supplemented by a unitarity-preserving
regularization.

Our regularization for self-energies~\cite{Astolfi:2011ju} is in essence equivalent to
the one suggested by Gromov and Mikhaylov~\cite{Gromov:2008fy},
who employ cutoffs, and for the sums consisting of heavy
$\omega^H$ and light $\omega^L$ modes use the prescription  \beq
\mathrm{Reg}\left(\sum \omega^H(n)+\omega^L(n)\right)\to \sum
\left(\omega^H(n)+\omega^L\left (\frac{n}{2}\right)\right).\eeq A
similar regularization has been applied in the one- and two-spin
BMN sector by Lipstein and Bandres~\mycite{Bandres:2009kw}. An
important finding of their paper is that the ``equal-frequency''
regularization leads to a linear divergence in the double-spin BMN
string for the algebraic curve result, whereas the
Gromov-Mikhaylov prescription yields all convergent results.

Let us mention here that in a very elucidating unpublished
Note\footnote{We thank Victor Mikhaylov for a clarifying
discussion on this note.} by Gromov, Mikhaylov and Vieira a very
general relation between the regularized one-loop self-energies
from algebraic curve/Bethe Ansatz on one side and worldsheet
string semiclassics on the other was derived. Gromov, Mikhaylov
and Vieira show in the Note that if equal frequency
(``world-sheet'') prescription is used then there are linear
divergences in self-energies and inconsistency between higher
charges of the integrable system as calculated from the discrete
Bethe Ansatz; that would mean that the Bethe Ansatz equations must
be modified in some way. On the other hand, if the unequal cutoff
(``algebraic curve'') prescription is used, then no linear
divergency arises and all charges are the same; Bethe Ansatz
remains valid in the form we know it. We emphasize here that the
result is stated in the Note by its authors as absolutely
universal, extending thus greatly the double-spin BMN string
obtained by  Bandres and Lipstein~\cite{Bandres:2009kw}; this
complies to our universal unitarity argument.

A special investigation is due on applicability of our
regularization prescription for the one-particle sector beyond the
near-BMN limit. This question is especially interesting with
regard to the generic GKP strings/twist-2 gauge operators.
Different subsectors of this sector include long and short
spinning folded strings, rotating circular strings, with $s$
large, very large or not very large (in each case a sophisticated
technical definition of ``largeness'' or ``smallness'' is
present). Certainly the physics is quite different from the BMN
case. The S-matrix argument we use here should be clarified in
adaptation to different sets of oscillations, since it was based
on a BMN-spectrum. The spectrum of GKP oscillations, as obtained
by Alday, Arutyunov and Bykov\mycite{Alday:2008ut}, contains of 6
bosons of zero mass, one boson of $\sqrt{2}$ mass, one mass 2
boson and 6 mass 1 fermions. It is not clear therefore how the
regularization (``unequal frequency'', ``algebraic curve'')
suggested by Astolfi, Grignani, Harmark and  Orselli
in~\cite{Astolfi:2011ju}, further argued for by Minahan and
Zarembo, explored by Gromov and Mikhaylov~\cite{Gromov:2008fy}
could be applied in its exact form\footnote{We specially thank
B.Basso for a discussion on this point.} in terms of mode numbers
on world-sheet. We conjecture that the unitarity argument will
work here as well, although the explicit form of the argument
based on the knowledge of certain pieces of the S-matrix would be
essentially different.

Thus we wish to draw once more the attention to the unitarity
issue in the regularization for generic sectors calling for
further research into this subject. A hint may be a very
interesting suggestion made by Gromov and Mikhaylov
in~\cite{Gromov:2008fy}. Namely the ``universal'' prescription is
to choose equal positions in the $x$-space for the excitations,
where the $x(u)$ algebraic curve coordinate. For all known cases,
such as $\ads_5$ and $\ads_4$ this automatically leads to the
correct mode structure and an ``algebraic-curve'' type
prescription which satisfies the unitarity condition and is
divergence-free. Thus a link, probably of a very general nature,
must be established between this simple scheme and the world-sheet
unitarity conservation.


\acknowledgments

A.Z. thanks Kolya Gromov and Dima Volin for their patience in
explaining him how the Bethe Ansatz works. We thank Benjamin
Basso, Diego Bombardelli, Constantin Candu, Sergey Frolov, Troels
Harmark, Kristan Jensen, Dmitry Kharzeev, Gregory Korchemsky,
Thomas Klose, Igor Klebanov, Marta Orselli, Tristan McLoughlin,
Juan Maldacena, Vitya Mikhaylov, Radu Roiban, Arkady Tseytlin,
Ismail Zahed and Alexander Zhiboedov  for stimulating
conversations. A.Z. thanks the Perimeter Institute, Princeton
University and Penn State University for hospitality, and the
Organizers of the August 2011 meeting on Strings and Integrability
at Perimeter for providing the fruitful atmosphere where an
essential part of this project was developed. Special thanks to
Fedor Levkovich-Maslyuk for a thorough reading of the manuscript
and comments on literature. This work was supported in part by the
MIUR-PRIN contract 2009-KHZKRX. The work of A.Z. is supported in
part by the RFBR grant 10-01-00836 supported by Ministry of
Education and Science of the Russian Federation under the contract
14.740.11.0081.



\addcontentsline{toc}{section}{References}


\providecommand{\href}[2]{#2}\begingroup\raggedright\endgroup

\end{document}